\documentclass[12pt,journal,onecolumn]{IEEEtran}
\usepackage{amsmath,amsfonts}
\usepackage{algorithmic}
\usepackage{algorithm}
\usepackage{array}
\usepackage[caption=false,font=normalsize,labelfont=sf,textfont=sf]{subfig}
\usepackage{textcomp}
\usepackage{stfloats}
\usepackage{url}
\usepackage{verbatim}
\usepackage{graphicx}
\usepackage{cite}
\usepackage{booktabs}
\usepackage{arydshln}
\usepackage{amssymb}
\usepackage{float}

\newcommand{\tr}{{\mathrm{Tr}}}

\newcommand{\gf}{{\mathbb{F}}}

\newcommand{\ord}{{\mathrm{ord}}}
\newtheorem{remark}{Remark}
\newtheorem{theorem}{Theorem}
\newtheorem{lemma}[theorem]{Lemma}
\newtheorem{Definition}[theorem]{Definition}
\newtheorem{proposition}[theorem]{Proposition}

\newtheorem{example}[theorem]{Example}

 \allowdisplaybreaks[3] 
\begin{document}

\title{Hybrid Character Sums From Vectorial Dual-Bent Functions and Asymptotically Optimal Complex Codebooks With Small Alphabet Sizes}

\author{Ziling Heng, Peng Wang and Chengju Li
        % <-this % stops a space
\thanks{Z. Heng and P. Wang are with the School of Science, Chang'an University, Xi'an 710064, China,  and also with the State Key Laboratory of Integrated Services Networks, Xidian University, Xi'an 710071, China (email: zilingheng@chd.edu.cn, wp20201115@163.com). C. Li is with the Shanghai Key Laboratory of Trustworthy Computing, East China Normal
University, Shanghai, 200062, China (email: cjli@sei.ecnu.edu.cn). \emph{(Corresponding author: Peng Wang)}}% <-this % stops a space
\thanks{Z. Heng's research was supported in part by the National Natural Science Foundation of China under Grant 12271059, in part by the Shaanxi Fundamental 
Science Research Project for Mathematics and Physics (Grant No. 23JSZ008), in part by the Foundation of State Key Laboratory of Integrated Services 
Networks under Grant ISN26-6 and in part by the Research Funds for the Interdisciplinary Projects, CHU, under Grant 300104240922. C. Li's research was supported by the National
Natural Science Foundation of China (T2322007, 12441103).} }

% \markboth{Journal of \LaTeX\ Class Files,~Vol.~1, No.~2, December~202}%

\IEEEpubid{0000--0000~\copyright~2023 IEEE}

\maketitle

\begin{abstract}
Hybrid character sums are an important class of exponential sums which have nice applications in coding theory and sequence design. 
Let $\gf_{p^m}$ be the finite field with $p^m$ elements for a prime $p$ and a positive integer $m$.
Let $V_n^{(p)}$ be an $n$-dimensional vector space over $\gf_p$ for a prime $p$.
In this paper, we study the hybrid character sums of the form 
\begin{eqnarray*}
\sum_{x \in V_n^{(p)}}\psi\left(F(x)\right)\chi_1\left(a x\right),
\end{eqnarray*}
where $F$ is a function from $V_n^{(p)}$ to $\gf_{p^m}$ and $a \in V_n^{(p)}$, $\psi$ is a  nontrivial multiplicative character of $\gf_{p^m}$ and $\chi_1$ is the canonical additive character of $V_n^{(p)}$.  If $F(x)$ is a vectorial dual-bent function and $a \in V_n^{(p)}\setminus \{0\}$, we  determine their complex modulus or explicit values  under certain conditions, which generalizes some known results as special cases. It is concluded that the hybrid character sums from vectorial dual-bent functions have very small complex modulus. As applications, three families of asymptotically optimal complex codebooks are constructed from vectorial dual-bent functions and their maximal cross-correlation amplitude are determined based on the hybrid character sums. The constructed codebooks have very small alphabet sizes, which enhances their appeal for implementation. Besides, all of the three families of codebooks have only two-valued or three-valued cross-correlation amplitudes. 
\end{abstract}


\section{Introduction}\label{sec1}
Hybrid character sums are an important class of exponential sums which have nice applications in coding theory and sequence design. 
In the following, we introduce the hybrid character sums, codebooks and the objectives of this paper. 
\subsection{Hybrid character sums}
Let $p$ be a prime, $q=p^m$ for a positive integer $m$ and $\gf_q$ be the finite field with $q$ elements. In \cite{Lidl}, a type of exponential sums, called the hybrid character sums (hybrid sums for short), was defined as
\begin{eqnarray}\label{eqn-1}
\sum_{x \in \gf_q}\psi(f(x))\lambda(g(x)),
\end{eqnarray}
where $f, g$ are functions over $\gf_q$, $\psi$ is a nontrivial multiplicative character and $\lambda$ is a nontrivial additive character of $\gf_q$. The hybrid sums can be viewed as a generalization of Gaussian sums. If $f,g\in \gf_q[x]$, the well-known Weil bound for the hybrid sums is given as follows.
\begin{theorem}\cite{Weil}\label{Weil bound}
Let $\psi$ be a nontrivial multiplicative character with order $M$ and $\lambda$ a nontrivial additive character of $\gf_q$. Define $\psi(0)=0$. Let $f,g \in \gf_q[x]$ such that $f$  has degree $d$ and $g$ has $s$ distinct roots in the algebraic closure of $\gf_q$ and $g(x)\neq c(h(x))^{M}$ for some $c \in \gf_q$ and $h(x) \in \gf_q[x]$. Then 
\begin{eqnarray*}
\left| \sum_{x \in \gf_q}\psi(f(x))\lambda(g(x)) \right| \leq (d+s-1)\sqrt{q}.
\end{eqnarray*}
\end{theorem}

Hybrid sums with small complex modulus are of particular  interest because they have many nice applications in coding theory, sequence design, combinatorics and many other fields\cite{Cao, Feng, Feng1, Zhou, Wang, WuX}. However, determining the complex modulus or explicit values of hybrid sums is difficult. To our knowledge, the known results on the complex modulus or explicit values of hybrid sums are presented as follows:
\begin{itemize}
\item{If $f(x)=x$ and $g(x)=x$, then the hybrid sums are known as Gaussian sums. The absolute values of Gaussian sums were given in \cite{Lidl}. The explicit values of Gaussian sums were determined for a few special cases \cite{Berndt,Lidl}.}
\item{If $q=p^{2e}, f(x)=\tr_{p^e/p}(x^{p^e+1}), g(x)=x$ and $\psi$ is the quadratic multiplicative character of $\gf_p$, then the explicit values of the hybrid sums were given in \cite{HongS}, where $\tr_{p^e/p}$ is the trace function from $\gf_{p^e}$ onto $\gf_p$.}
\item{If $q=p^{2e}$, $e'\mid e, f(x)=\tr_{p^e/p^{e'}}(x^{p^e+1}), g(x)=x$ and $\psi$ is the quadratic multiplicative character of $\gf_{p^{e'}}$, then the explicit values of the hybrid sums were studied in \cite{WuX}, where $\tr_{p^e/p^{e'}}$ is the trace function from $\gf_{p^e}$ onto $\gf_{p^{e'}}$.}
\item{If $f(x)=x$ and $g(zx)-g(x)$ permutes $\gf_q$ for any $z\in \gf_q\backslash \{1\}$ with $g(0)=0$, then the complex modulus of the hybrid sums were determined in \cite{Cao}.}
\item{If $f(x) \in \mathcal{RF}$, $g(x)=ax$ for $a\in \gf_q^*$ and $\psi$ is a nontrivial multiplicative character of $\gf_p$, then the complex modulus or the explicit values of the hybrid character sums were calculated in \cite{Heng}, where $\mathcal{RF}$ is a  set of weakly regular bent functions satisfying certain conditions.}
\item If $q=p^m$ for $2e\mid m$, $f(x)=\tr_{p^{\frac{m}{2}}/p^e}(x^{\frac{m}{2}+1})$, $g(x)=ax$ for $a\in \gf_q^*$ and $\psi$ is a nontrivial multiplicative character of $\gf_{p^e}$, then the complex modulus or the explicit values of the hybrid character sums were studied in \cite{Heng}.
\item If $q=p^m$ for $e\mid m$, $f(x)=\tr_{q/p^e}(x^2)$, $g(x)=ax$ for $a\in \gf_q^*$ and $\psi$ is a nontrivial multiplicative character of $\gf_{p^e}$, then the complex modulus or the explicit values  of the hybrid sums were determined in \cite{Heng}.
\end{itemize}
The hybrid sums in these known cases were used to construct asymptotically optimal codebooks \cite{Zhou, HongS, WuX, Heng} and approximately symmetric informationally complete positive operator-valued measures \cite{Cao}. The maximum complex modulus of these known hybrid sums are about $\sqrt{q}$.

\subsection{Codebooks}
An $(N,K)$ complex codebook $\mathcal{C}=\{\mathbf{c}_l\}_{l=0}^{N-1}$ is defined as a set of $N$ distinct unit-norm complex row vectors $\mathbf{c}_l \in \mathbb{C}^{K}$ of length $K$ over an alphabet $A$, where the size of $A$ is referred to as the alphabet size of $\mathcal{C}$.
For an $(N,K)$ codebook $\mathcal{C}$, the maximum magnitude of the inner products between a pair of distinct vectors in $\mathcal{C}$ is defined by
\begin{eqnarray*}
I_{\max}(\mathcal{C})=\max_{0 \leq l \neq k \leq N-1}\mid\mathbf{c}_l\mathbf{c}_k^{H}\mid,
\end{eqnarray*}
where $\mathbf{c}_k^{H}$ denotes the conjugate transpose of $\mathbf{c}_k$.  $I_{\max}(\mathcal{C})$ is also called the maximal cross-correlation amplitude of $\mathcal{C}$. 
For practical applications, it is desirable to construct a codebook minimizing the maximum magnitude.
According to \cite{Hochwald}, minimizing the maximal cross-correlation amplitude of a codebook is equivalent to minimizing the block error probability in the context of unitary space-time modulations.

For fixed $K$, we hope to construct an $(N,K)$ codebook $\mathcal{C}$ with $N$ as large as possible and $I_{\max}(\mathcal{C})$ as small as possible.
Besides, the alphabet size of $\mathcal{C}$ is required to be small to enhance its appeal for implementation.
However, there exists a tradeoff among the parameters of a codebook. The well-known Welch bound on $I_{\max}(\mathcal{C})$ is given as follows.
\begin{lemma}[Welch bound]\label{welch}
Let $\mathcal{C}$ be an $(N,K)$ codebook for $N\geq K$. Then 
\begin{eqnarray}\label{welch1}
I_{\max}(\mathcal{C})\geq I_{W}:=\sqrt{\frac{N-K}{K(N-1)}},
\end{eqnarray}
where the equality holds if and only if 
\begin{eqnarray*}
\mid\mathbf{c}_l\mathbf{c}_k^{H}\mid=\sqrt{\frac{N-K}{K(N-1)}}
\end{eqnarray*}
for all pairs of $(l,k)$ with $0\leq l\neq k\leq N-1$.
\end{lemma}

If an $(N,K)$ codebook $\mathcal{C}$ achieves the Welch bound with equality, then $\mathcal{C}$ is called a maximum-Welch-bound-equality (MWBE) codebook \cite{Sarwate} or an equiangular tight frame \cite{Kovacevic}. MWBE codebooks have widely applications in many areas including compressed sensing\cite{Candes}, quantum computing \cite{Renes}, synchronous
code-division multiple access systems \cite{Massey}, packing \cite{Conway}\cite{Strohmer}, coding theory\cite{Calderbank}\cite{Delsarte} and so on.
MWBE codebooks with $N=K$ or $N=K-1$ are said to be trivial ones as they
can be easily constructed from the discrete Fourier transform matrix. However, constructing a nontrivial MWBE codebook is not easy. There are only a few families
of MWBE codebooks reported in the literature. We list them as follows:
\begin{itemize}
\item{$(N,K)$ MWBE codebooks from cyclic difference sets in the residual class additive group $(\mathbb{Z}_{N},+)$ of modular $N$ or the additive group of finite fields or general Abelian groups\cite{Ding},\cite{Ding1},\cite{XiaP};}
\item{$(N,K)$ MWBE codebooks from $(2,k,v)$-Steiner system\cite{Fickus};}
\item{$(N,K)$MWBE codebooks from conference matrices with $N=2K$ \cite{Conway}\cite{Strohmer};}
\item{$(N,K)$MWBE codebooks from graph theory and finite geometries\cite{Fickus1}\cite{Rahimi}.}
\end{itemize}

If $N>\frac{K(K+1)}{2}$ for a real codebook and $N>K^2$ for a complex codebook, it is known that the Welch bound for $I_{\max}(\mathcal{C})$ of an $(N,K)$ codebook $\mathcal{C}$ is not tight. Then the Levenshtein bound, denoted by $I_{L}$, was proposed in \cite{Levens} to improve the Welch bound in these cases. 
Optimal codebooks with respect to the Levenshtein bound were reported in \cite{Calderbank} \cite{Ding2} \cite{Heng1} \cite{XiangC} \cite{Zhou2}.

As was pointed out in \cite{Sarwate}, constructing an optimal codebook is very difficult. Since the known classes of optimal codebooks achieving the Welch bound or the Levenshtein bound have very restrictive $N$ and $K$, constructing asymptotically optimal codebooks with flexible and new parameters has been a hot topic in recent years.
For an $(N,K)$ codebook $\mathcal{C}$, if $\lim_{K\rightarrow +\infty}\frac{I_{\max}(\mathcal{C})}{I_W}=1$ (resp. $\lim_{K\rightarrow +\infty}\frac{I_{\max}(\mathcal{C})}{I_L}=1$),
then  $\mathcal{C}$ is said  to be \emph{asymptotically optimal} with respect to the Welch bound (resp. Levenshtein bound). In \cite{HongS} \cite{NY}, some binary sequences were used to construct codebooks that asymptotically achieve the Welch bound. In \cite{Ding3} \cite{LiC} \cite{Zhang} \cite{Zhang1},  some families of codebooks that asymptotically achieve the Welch bound were constructed based on some almost difference sets or partial difference sets. In\cite{Heng1,Heng2,Heng3,Zhou}, some exponential sums including the Jacobi sums and Gaussian sums were utilized to construct asymptotically optimal codebooks with respect to the Levenshtein bound. In \cite{XiangC}, codebooks asymptotically achieving the Levenshtein bound were obtained from binary codes.

\subsection{The objectives of this paper}
Since the hybrid character sums with small complex modulus are very essential in coding theory and sequence designs, it motivates us to further study hybrid character sums as well as their nice applications. Let $V_n^{(p)}$ be an $n$-dimensional vector space over $\gf_p$ for a prime $p$.
In this paper, we generalize the hybrid character sums from the form (\ref{eqn-1}) to
\begin{eqnarray}\label{form2}
\sum_{x \in V_n^{(p)}}\psi\left(F(x)\right)\chi_1\left(G(x)\right),
\end{eqnarray}
where $F$ is a function from $V_n^{(p)}$ to $\gf_{q}$ and $G(x)$ is a function from $V_n^{(p)}$ to  $V_n^{(p)}$, $\psi$ is a  nontrivial multiplicative character of $\gf_{p^m}$ and $\chi_1$ is the canonical additive character of $V_n^{(p)}$.  If $F(x)$ is a vectorial dual-bent function under certain conditions and $G(x)=ax$ for $a \in V_n^{(p)}\setminus \{0\}$, we  determine their complex modulus or explicit values, which generalizes some known results in \cite{Heng} as special cases. It is concluded that the hybrid character sums from vectorial dual-bent functions have very small complex modulus. As applications, three families of asymptotically optimal complex codebooks are constructed from vectorial dual-bent functions and their maximal cross-correlation amplitude are determined based on the hybrid character sums. The constructed codebooks have very small alphabet size, which enhances their appeal for implementation. Besides, all of the three families of codebooks have only two-valued or three-valued cross-correlation amplitudes. Compared with the codebooks in \cite{Heng} \cite{WuX}, the codebooks constructed in this paper have more flexible parameters.

\section{Preliminaries}\label{sec2}
In this section, we introduce some notations and auxiliary results which will be used to give our main results later.
\subsection{Notations}
We adopt the following notations throughout this paper.
\begin{itemize}
\item{$q=p^m$, where $p$ is a prime and $m$ is a positive integer}.
\item $\tr_{p^n/p^t}$ denotes the trace function from $\gf_{p^n}$ to $\gf_{p^t}$ for $t\mid n$.
\item{$\epsilon:=\sqrt{(-1)^{\frac{p-1}{2}}}$}.
\item{$\zeta_{p}=e^{\frac{2\pi\sqrt{-1}}{p}}$ is a complex primitive $p$-th root of unity}.
\item{$\gf_q$ is a finite field with $q$ elements and $\gf_q^{*}=\gf_q\setminus\{0\}$}.
\item{$\gf_q^{n}$ is the vector space of $n$-tuples over $\gf_q$}.
\item{$V_n^{(p)}$ is an $n$-dimensional vector space over $\gf_p$}.
\item{$\langle,\rangle_n$ denotes a (non-degenerate) inner product of $V_n^{(p)}$. In this paper, when $\langle,\rangle_n=\gf_q^{n}$, let $\langle \textbf{a},\textbf{b}\rangle_n := \textbf{a}\cdot \textbf{b}= \sum_{i=1}^{n}a_{i}b_{i}$ called the \emph{standard inner product}, where $\textbf{a}=(a_{1},\ldots,a_{n}),\textbf{b}=(b_{1},\ldots,b_{n})\in \gf_p^{n}$; when $V_{n}^{(p)}=\gf_{p^n}$, let $\langle a,b\rangle_n:=\tr_{p^n/p}(ab)$ called the \emph{trace inner product}, where $a,b\in\gf_{p^n}$; when $V_n^{(p)}=V_{n_1}^{(p)}\times\cdots\times V_{n_s}^{(p)}$, let $\langle \textbf{a},\textbf{b}\rangle_n:= \sum_{i=1}^{s}\langle a_i,b_i\rangle_n$, where $\textbf{a}=(a_1,\ldots, a_s), \textbf{b}=(b_1,\cdots,b_s)\in V_{n}^{(p)}$.}
\item{If $V_n^{(p)}=\gf_{p^{n_1}}\times\gf_{p^{n_2}}\times\cdots\times\gf_{p^{n_s}}$ and $\textbf{x}\in V_{n}^{(p)}$, denote $\textbf{x}=(x_1,\ldots,x_s)$, where $x_j\in \gf_{p^{n_j}}, 1\leq j\leq s.$}  
\item{For any set $A\subseteq V_n^{(p)}$ and $a\in V_n^{(p)}$, let $\chi_{a}(A):=\sum_{x\in A}\chi_{a}(x)$, where $\chi_{a}$ is the additive character defined by $\chi_a(x)=\zeta_{p}^{\langle a,x\rangle_n}$}.
\item{$\lambda_y, y\in \gf_{q}$, is an additive character of $\gf_{q}$.}
\item $\eta_m$ denotes the quadratic multiplicative characters of $\gf_{q}$.     
\end{itemize}
\subsection{Characters and Gaussian sums over finite fields}
The \emph{trace function} from $\gf_{q}$ onto $\gf_p$ is defined as 
\begin{eqnarray*}
\tr_{q/p}(x)=x+x^p+x^{p^2}+\cdots+x^{p^{m-1}}, x\in\gf_{q}.
\end{eqnarray*}
An \emph{additive character} $\lambda$ of $\gf_{q}$ is defined as a homomorphism from $\gf_{q}$ onto the multiplicative group of $p$-th roots of complex unity satisfying $\lambda(x_1+x_2)=\lambda(x_1)\lambda(x_2)$ for all $x_1, x_2\in \gf_{q}$. The conjugate $\overline{\lambda}$ of $\lambda$ is defined by $\overline{\lambda}(x)=\overline{\lambda(x)}=\lambda(-x)$ for all $x\in \gf_{q}$. For any $a\in \gf_{q}$, an additive character of $\gf_q$ can be given by 
\begin{eqnarray*}
\lambda_{a}(x)=\zeta_p^{\tr_{q/p}(ax)}, x\in \gf_{q}.
\end{eqnarray*}
All the additive characters of $\gf_q$ form a character group $\widehat{\gf}_{q}:=\{\lambda_a: a \in \gf_{q}\}$. Specially, $\lambda_{0}$ and $\lambda_{1}$ are called the \emph{trivial} and \emph{canonical additive characters} of $\gf_{q}$, respectively. 
In \cite{Lidl}, the orthogonality relation for additive characters of $\gf_{q}$ was given as
\begin{eqnarray*}
\sum_{x\in\gf_{q}}\lambda_{a}(x)=\sum_{x\in \gf_{q}}\lambda_1(ax)=\left\{
\begin{array}{ll}
q  &   \mbox{if $a=0$},\\
0    &   \mbox{if $a\in \gf_{q}^*$}.
\end{array} \right.
\end{eqnarray*}

Let $\gf_{q}^{*}=\langle\beta\rangle$. The \emph{multiplicative character} of $\gf_{q}$ is defined as the homomorphic function $\psi$ from $\gf_{q}$ onto the multiplicative group of $(q-1)$-th root of complex unity such that $\psi(xy)=\psi(x)\psi(y)$ for $x,y\in \gf_q^*$. For each $j=0, 1, \cdots, p^m-2$, then the following function $\psi_j$ such that
\begin{eqnarray*}
\psi_j(\beta^s)=\zeta_{q-1}^{sj}, 0\leq s\leq q-2,
\end{eqnarray*}
is an multiplicative character of $\gf_{q}$. In fact, the set $\widehat{\gf}_{q}:=\{\psi_j: j=0, 1, \cdots, q-2\}$ consists of all the multiplicative characters of $\gf_{q}$. For any two multiplicative characters $\psi_i$ and $\psi_j$, we define $\psi_i\psi_j(x)=\psi_i(x)\psi_j(x)$, $x\in \gf_{q}^{*}$. The conjugate $\overline{\psi}$ of a multiplicative character $\psi$ is defined by $\overline{\psi}(x)=\overline{\psi(x)}=\psi(x^{-1})$.
 If $q$ is odd, then $\eta_m:=\psi_{\frac{q-1}{2}}$ is called the \emph{quadratic multiplicative character} of $\gf_{q}$. Besides, $\psi_{0}$ is called the trivial multiplicative character of $\gf_{q}$. In \cite{Lidl}, the orthogonality relation for multiplicative characters of $\gf_{q}$  was given by 
\begin{eqnarray*}
\sum_{x\in \gf_{q}^{*}}\psi_j(x)=\left\{
\begin{array}{ll}
q-1  &   \mbox{if $j=0$},\\
0    &   \mbox{if $0<j\leq q-2$}.
\end{array} \right.
\end{eqnarray*}
In addition, we extend the definition of the multiplicative character $\psi_j$ of $\gf_{q}$ by 
\begin{eqnarray*}
\psi_j(0)=\left\{
\begin{array}{ll}
1  &   \mbox{if $j=0$},\\
0    &   \mbox{if $0<j\leq q-2$}.
\end{array} \right.
\end{eqnarray*}

For an additive character $\lambda$ and a multiplicative character $\psi$ of $\gf_{q}$, the \emph{Gaussian sum} $G(\psi,\lambda)$ over $\gf_{q}$ is defined by 
\begin{eqnarray*}
G(\psi,\lambda)=\sum_{x\in \gf_{q}^{*}}\psi(x)\lambda(x).
\end{eqnarray*}
Specially, $G(\eta_m,\lambda)$ is called the quadratic Gaussian sum for $\lambda\neq\lambda_{0}$.
Next, we recall the properties of Gaussian sums.
\begin{lemma}\cite{Lidl}\label{quadGuasssum1}
The properties of Gaussian sums are given as follows:
\begin{enumerate}
\item $G(\psi,\lambda_{ab})=\overline{\psi(a)}G(\psi,\lambda_{b})$ for $a\in\gf_{q}^{*}$ and $b\in\gf_{q}$;
\item $G(\psi,\overline{\lambda})=\psi(-1)G(\psi,\lambda)$;
\item $G(\overline{\psi},\lambda)=\psi(-1)\overline{G(\psi,\lambda)}$.
\end{enumerate}
\end{lemma}

\begin{lemma}\cite{Lidl}\label{quadGuasssum2}
Let $\psi$ be a multiplicative character and $\lambda$ be an additive character of $\gf_{q}$. Then 
\begin{eqnarray*}
G(\psi,\lambda)=\left\{
\begin{array}{lll}
q-1  &   \mbox{if $\psi=\psi_0$, $\lambda=\lambda_0$},\\
-1    &   \mbox{if $\psi=\psi_0$, $\lambda\neq\lambda_0$},\\
0    &   \mbox{if $\psi\neq\psi_0$, $\lambda=\lambda_0$}.\\
\end{array} \right.
\end{eqnarray*}
If $\psi\neq\psi_0$ and $\lambda\neq\lambda_0$, then $|G(\psi,\lambda)|=\sqrt{q}$.
\end{lemma}

Determining the values of Gaussian sums is difficult in general. The explicit values of Gaussian sums are known only in some special cases.
\begin{lemma}[The quardratic Gaussian sums]\cite{Lidl}\label{quadGuasssum3}
Let $p$ be an odd prime. Let $\lambda_1$ be the canonical additive character of $\gf_{q}$. Then 
\begin{eqnarray*}
G(\eta_m,\lambda_1)&=&(-1)^{m-1}(\sqrt{-1})^{(\frac{p-1}{2})^2m}\sqrt{q}\\
 &=&\left\{
\begin{array}{lll}
(-1)^{m-1}\sqrt{q}    &   \mbox{for }p\equiv 1\pmod{4},\\
(-1)^{m-1}(\sqrt{-1})^{m}\sqrt{q}    &   \mbox{for }p\equiv 3\pmod{4}.
\end{array} \right.
\end{eqnarray*}
\end{lemma}

% \begin{lemma}[The semi-primitive case Gaussian sums]\cite{Berndt}\label{quadGuasssum4}
% Let $r$ be a prime power and $\psi$ be a multiplicative character with order $N$ of $\gf_r$ and $N>2$. Let $\lambda_1$ be the canonical additive character of $\gf_r$. Let $\ell$ be the least positive integer satisfy $p^\ell\equiv-1 \mod N$ and $r=p^{2\ell m}$ for some positive integer $m$. If $1\leq i \leq N-1$, then
% \begin{eqnarray*}
% G(\psi^{i},\lambda_1)=\left\{
% \begin{array}{ll}
% (-1)^i\sqrt{r}  &   \mbox{if $N$ is even,~$p, m$ and $\frac{p^\ell+1}{N}$ are odd},\\
% (-1)^{m-1}\sqrt{r}    &   \mbox{otherwise.}\\
% \end{array} \right.
% \end{eqnarray*}
% \end{lemma}

In the following, we give the Fourier expansion of a multiplicative character in terms of the additive characters  of $\gf_{q}$ with Gaussian sums appearing as Fourier coefficients.
\begin{lemma}\cite{Heng}\label{Fourier1}
Let $\psi$ be a multiplicative character of $\gf_{q}$, then
\begin{eqnarray*}
\psi(x)=\frac{1}{q}\sum_{\lambda\in\widehat{\gf}_{q}}G(\psi,\overline{\lambda})\lambda(x),\ x\in \gf_{q}.
\end{eqnarray*}
\end{lemma}

The Fourier expansion of  an additive character in terms of the multiplicative characters of $\gf_{q}$ with Gaussian sums appearing as Fourier coefficients is given as follows.
\begin{lemma}\cite{Lidl}\label{Fourier2}
Let $\lambda$ be an additive character of $\gf_{q}$, then
\begin{eqnarray*}
\lambda(x)=\frac{1}{q-1}\sum_{\psi \in \widehat{\gf}_{q}^*}G(\overline{\psi},\lambda)\psi(x),\ x \in \gf_{q}^{*}. 
\end{eqnarray*}
\end{lemma}

\subsection{Partial Hadamard codebooks}
In the following, we introduce the Hadamard matrix and partial Hadamard codebooks.
Let $\gf_{p^n}^*=\langle \alpha \rangle.$ Denote by $s(t)$ the $p$-ary $m$-sequence of period $p^n-1$, i.e., 
\begin{eqnarray*}
s(t)=\tr_{p^n/p}(\alpha^t),
\end{eqnarray*}
where $t=0,1,\cdots, p^n-2$. The complex Hadamard matrix is defined based on the $p$-ary $m$-sequence as follows.
\begin{Definition}\label{def1}
Let $\mathbf{H}=[h_{i,j}]_{0\leq i,j\leq p^n-1}$ be the $p^n\times p^n$ $p$-ary Hadamard matrix  defined as
\begin{eqnarray*}
h_{i,j}=\left\{
\begin{array}{ll}
1  &   \mbox{if}~ i=0 ~ \mbox{or} ~j=0, \\
\zeta_{p}^{\tr_{p^n/p}(\alpha^{i+j-2})}   &    \mbox{otherwise}.\\
\end{array} \right.
\end{eqnarray*}
where $0\leq i, j\leq p^n-1.$
\end{Definition}

For a binary sequence $\mathbf{a}=(a_0,a_1,\cdots,a_{N-1})$ of length $N$, its \emph{support} $D$ is defined as $$D=\{j:0\leq j \leq N-1, a_j=1\}.$$ 
The following lemma gives the construction of partial Hadamard codebooks from the Hadamard matrix, whose maximal cross-correlation amplitude is related to the Hadamard transform of the row selection sequence.
\begin{lemma}\cite{NY}\label{bina}
Define a binary row selection sequence $\mathbf{a}=(a_0,a_1,\cdots,a_{N-1})$ with support $D=\{d_0,d_1,\cdots,d_{K-1}\}.$ Let $\mathcal{C}_{H}(\mathbf{a})=\{\mathbf{c}_0,\mathbf{c}_1,\ldots,\mathbf{c}_{N-1}\}$ be an $(N,K)$ partial Hadamard codebook whose code vectors are the columns of the submatrix obtianed by selecting $K$ rows corresponding to $D$ from the $p^n\times p^n$ Hadamard matrix $\mathbf{H}=[h_{i,j}]_{0\leq i,j\leq p^n-1}$. For different integers $i,j$, $0\leq i,j \leq N-1$, let $I_{i,j}(\mathcal{C}_{H}(\mathbf{a}))$ be the magnitude of the inner product between the code vectors $\mathbf{c}_i$ and $\mathbf{c}_j$. Then, for any pair $(i,j),i\neq j$, there exists $l$ for $1\leq l\leq N-1$ satisfying 
\begin{eqnarray*}
I_{i,j}(\mathcal{C_{H}(\mathbf{a})})=\frac{1}{2K}| \widehat{a_l}|,
\end{eqnarray*}
and the number of pairs $(i,j)$ corresponding to each $l$ equals $N$, where $\widehat{a_l}:=\sum_{k=0}^{N-1}(-1)^{a_k}h_{k,l}$ is the $l$-th element of the Hadamard transform of the row selection
sequence $\mathbf{a}$.
\end{lemma}

\subsection{Vectorial dual-bent functions}
A function $F: V_{n}^{(p)}\rightarrow V_{m}^{(p)}$ is called a vectorial $p$-ary function.
If $m=1$, the vectorial $p$-ary function is the usual $p$-ary function.

For a $p$-ary function $f: V_{n}^{(p)}\rightarrow \gf_p$, its Walsh transform is defined by
\begin{eqnarray*}
W_{f}(a)=\sum_{x \in V_{n}^{(p)}}\zeta_{p}^{f(x)-\langle a,x\rangle_n},\ a\in V_{n}^{(p)}.
\end{eqnarray*}
The inverse Walsh transform of $f(x)$ is defined as
\begin{eqnarray*}
\zeta_{p}^{f(x)}=\frac{1}{p^n}\sum_{a \in V_{n}^{(p)}}W_{f}(a)\zeta_{p}^{\langle a, x\rangle_n},\ x \in V_{n}^{(p)}.
\end{eqnarray*}
If $| W_{f}(a)|=p^{\frac{n}{2}}$ for all $a \in V_{n}^{(p)}$, then $f$ is called a $p$-ary\emph{ bent function}. 
For a $p$-ary bent function $f$, if $p=2$, then $W_{f}(a)=2^{n/2}(-1)^{f^*(a)}$; if $p$ is an odd prime, then
\begin{eqnarray*}
W_{f}(a)=\left\{
\begin{array}{ll}
\pm p^{\frac{n}{2}}\zeta_{p}^{f^{*}(a)}  &   \mbox{if  $p \equiv 1 \pmod 4$ or $n$ is even},\\
\pm \sqrt{-1}p^{\frac{n}{2}}\zeta_{p}^{f^{*}(a)}   &   \mbox{if $p \equiv 3 \pmod 4 $ and $n$ is odd},\\
\end{array} \right.
\end{eqnarray*}
where $f^*: V_{n}^{(p)}\rightarrow \gf_p$  is the \emph{dual} of $f$ \cite{Tang}. A $p$-ary function $f$ is said to be \emph{regular bent} if $W_{f}(a)=p^{\frac{n}{2}}\zeta_{p}^{f^{*}(a)}$ for any $a \in V_n^{(p)}$, and \emph{weakly regular bent} if $W_{f}(a)=\varepsilon_{f}p^{\frac{n}{2}}\zeta_{p}^{f^{*}(a)}$ for any $a \in V_n^{(p)}$, where $\varepsilon_{f} \in \{\pm1,\pm\sqrt{-1}\}$ is referred to as the \emph{sign of the walsh transform of} $f$. Otherwise, the $p$-ary bent function $f$ is said to be \emph{non-weakly regular}. 
It is known from \cite{Tang} that the dual $f^{*}$ of a $p$-ary weakly regular bent function $f$ is also a weakly regular bent function such that
\begin{eqnarray*}
 (f^*)^*(x)=f(-x),\ \varepsilon_{f^*}=\varepsilon_{f}^{-1}.
\end{eqnarray*}

A vectorial $p$-ary function $F:V_{n}^{(p)}\rightarrow V_{m}^{(p)}$ is said to be \emph{vectorial bent} if the component function $F_c$ defined as $F_{c}(x)=\langle c, F(x)\rangle_{m}$ is a $p$-ary bent function \cite{Coesmelioglu1}. It is clear that every $p$-ary bent function is vectorial bent. For a vectorial $p$-ary bent function $F: V_{n}^{(p)}\rightarrow V_{m}^{(p)}$ and any $c \in V_{m}^{(p)}\setminus\{0\}$, if there exists a vectorial bent function $G:V_{n}^{(p)}\rightarrow V_{m}^{(p)}$ satisfying $(F_{c})^{*}=G_{\sigma(c)}$, 
then $F$ is said to be \emph{vectorial dual-bent}, where $(F_{c})^{*}$ is the dual of $F_c$ and $\sigma$ is some permutation over $c \in V_{m}^{(p)}\setminus\{0\}$. The vectorial bent function $G$ is referred to as  a \emph{vectorial dual} of $F$ which is denoted by $F^*$.

A $p$-ary function $f:V_{n}^{(p)}\rightarrow \gf_p$ is said to be of $l$-form if $f(ax)=a^lf(x)$ for any $a \in \gf_p^*$, $x \in V_{n}^{(p)}$ and some integer $l$. In \cite{Tang}, 
 the set consisting of all weakly regular bent functions $f:V_{n}^{(p)}\rightarrow \gf_p$ of $l$-form such that $f(\textbf{0})=0$ and $\gcd(l-1,p-1)=1$ is denoted by $\mathcal{RF}$.
\begin{proposition}\cite{Coesmelioglu}
Let $f: V_{n}^{(p)}\rightarrow \gf_p$ be a bent function belonging to $\mathcal{RF}$. Then $f$ is a vectorial dual-bent function with $(cf)^*=c^{1-d}f^*$, $c \in \gf_p^*$, and $\varepsilon_{cf}=\varepsilon_{f}$ if $n$ is even, $\varepsilon_{cf}=\varepsilon_{f}\eta_{1}(c)$ if $n$ is odd, where $(l-1)(d-1)\equiv 1\mod {(p-1)}$.
\end{proposition}

\section{Evaluation of hybrid character sums from vectorial dual-bent functions}
In this section, we study the hybrid character sums of the following type:
\begin{eqnarray}\label{sum-1}
\sum_{x \in V_n^{(p)}}\psi\left(F(x)\right)\chi_1\left(a x\right),
\end{eqnarray}
where $F$ is a function from $V_n^{(p)}$ to $\gf_{p^m}$ and $a \in V_n^{(p)}$, $\psi$ is a  nontrivial multiplicative character of $\gf_{p^m}$ and $\chi_1$ is the canonical additive character of $V_n^{(p)}$.  If $F(x)$ is a vectorial dual-bent function under certain conditions and $a \in V_n^{(p)}\setminus \{0\}$, we will determine their complex modulus or explicit values.

\subsection{The case that $F(x)$ satisfies Condition \uppercase\expandafter{\romannumeral1}}
In this subsection, we consider the case that $F(x)$ satisfies Condition \uppercase\expandafter{\romannumeral1} as follows.

\textbf{Condition \uppercase\expandafter{\romannumeral1}}: Let $p$ be an odd prime, $n, n_{j}, 1\leq j\leq s$, $m$, $t$ be positive integers where $n=\sum_{j=1}^{s}n_{j}$, $2\mid n$, $t\mid n_{j}$,  $t\mid m$, $m\leq\frac{n}{2}$ and $V_n^{(p)}=\gf_{p^{n_1}}\times \gf_{p^{n_2}}\times\cdots\times \gf_{p^{n_s}}$. Let $F: V_{n}^{(p)}\rightarrow \gf_{p^m}$ be a vectorial dual-bent function satisfying that
 \begin{itemize}
\item{there is a vectorial dual $F^*$ of $F$ satisfying ${F_c}^{*}=(F^*)_{c^{1-d}}$}, $c \in \gf_{p^m}^{*}$, where $\gcd(d-1,p^m-1)=1$;
\item{$F(bx)=b^{l}F(x)$, $b \in \gf_{p^t}^{*}$, $x \in V_{n}^{(p)}$ and $F(0)=0$, where $(l-1)(d-1)\equiv 1 \mod ({p^m-1})$};
\item{all component functions $F_c, c \in \gf_{p^m}^{*}$, are weakly regular with $\varepsilon_{F_{c}}=\varepsilon$, $c \in \gf_{p^m}^{*}$, where $\varepsilon \in \{\pm1\}$ is a constant}.
\end{itemize}

The following lemma indicates that the vectorial dual of a vectorial dual-bent function satisfying Condition \uppercase\expandafter{\romannumeral1} is also a vectorial dual-bent function satisfying Condition \uppercase\expandafter{\romannumeral1}.
\begin{lemma}\cite{WangJ2}\label{lem8}
Let $F$ be a vectorial dual-bent function satisfying Condition \uppercase\expandafter{\romannumeral1}. Then the vectorial dual $F^{*}$ with $(F_c)^{*}=(F^*)_{c^{1-d}}$, $c\in \gf_{p^m}^*$, is a vectorial dual-bent function such that $((F^*)_c)^*=F_{c^{1-l}}$, $c \in \gf_{p^m}^*$, $F^*(ax)=a^{d}F^*(x)$, $a \in \gf_{p^t}^*$, $F^*(0)=0$, and all component functions $(F^*)_c, c \in \gf_{p^m}^*$, are weakly regular with $\varepsilon_{(F^*)_c}=\varepsilon.$
\end{lemma}

In the following, based on the results from \cite{Coesmelioglu1}\cite{WangJ}\cite{WangJ1}, we  present some explicit classes of vectorial dual-bent functions $F: V_{n}^{(p)}\rightarrow \gf_{p^m}$ satisfying Condition \uppercase\expandafter{\romannumeral1}.
\begin{itemize}
\item{Let $p$ be an odd prime, $m$,\ $n'$,\ $t$,\ $u$ be positive integers with $t\mid m, m\mid n', m\neq n', \gcd(u,p^{n'}-1)=1$. Let $e \in \gf_{p^{n'}}^{*}$. Define $F: \gf_{p^{n'}}\times\gf_{p^{n'}}\rightarrow \gf_{p^m}$ as}
 \begin{eqnarray*}
 F(x_1,x_2)=\tr_{p^{n'}/p^m}(e x_1x_2^{u}).
 \end{eqnarray*}
 Then $F$ is a vectorial dual-bent function satisfying Condition \uppercase\expandafter{\romannumeral1} with $l=1+u, d=1+u', \varepsilon=1$, where $uu'=1 \mod(p^{n'}-1)$ \cite{Coesmelioglu1}.
 \item{Let $p$ be an odd prime, $t$, $m$, $s$ be positive integers with $t\mid m$, $2\mid s$. Based on the results in \cite{WangJ} \cite{WangJ1}, all the non-degenerate quadratic forms $F$ from $\gf_{p^m}^{s}(\gf_{p^{ms}})$ to $\gf_{p^m}$ are vectorial dual-bent functions satisfying Condition \uppercase\expandafter{\romannumeral1} with $l=d=2$. The followings are a list of such functions.}
\begin{enumerate}
\item{Let $m,\ n,\ t$ be positive integers with $t\mid m$, even $\frac{n}{m}$, $e \in \gf_{p^n}^{*}$. Define $F: \gf_{p^n}\rightarrow \gf_{p^m}$ as}
     \begin{eqnarray}\label{Tr(x^2)}
     F(x)=\tr_{p^n/p^m}(e x^2).
     \end{eqnarray}
     Then $F$ is a vectorial dual-bent function satisfying Condition \uppercase\expandafter{\romannumeral1} with $l=d=2, \varepsilon=-\epsilon^n\eta_{n}(e).$
\item{Let $m,\ n,\ t$ be positive integers with $t\mid m$, $s$ being even, $\alpha_i \in \gf_{p^m}^{*}$, $1\leq i \leq s$. Define $F: \gf_{p^m}^{s}\rightarrow \gf_{p^m}$ by}
    \begin{eqnarray*}
    F(x_1, x_2,\ldots, x_s)=\sum_{i=1}^{s}\alpha_{i}x_{i}^2.
    \end{eqnarray*}
    Then $F$ is a vectorial dual-bent function satisfying Condition \uppercase\expandafter{\romannumeral1} with $l=d=2,\varepsilon=\epsilon^{ms}\eta_{m}(\alpha_1\cdots\alpha_s)$.
\item{Let $m, n, t$ be positive integers with $t\mid m,$ $\frac{n}{m}$ being even, $ e \in \gf_{p^n}^{*}$. Define $F: \gf_{p^n}\rightarrow \gf_{p^m}$ by}
    \begin{eqnarray*}
    F(x)=\tr_{p^{\frac{n}{2}}/p^m}(e x^{p^{\frac{n}{2}}+1}).
    \end{eqnarray*}
     Then $F$ is a vectorial dual-bent function satisfying Condition \uppercase\expandafter{\romannumeral1} with $l=d=2, \varepsilon=-1.$
\end{enumerate}
\end{itemize}

If $\psi=\eta_m$ (i.e. the quadratic multiplicative character of $\gf_{p^m}$) and $F$ is a vectorial dual-bent function satisfying Condition \uppercase\expandafter{\romannumeral1}, we denote the hybrid character sums in (\ref{sum-1}) as
$$S_1=\sum_{x \in V_n^{(p)}}\eta_m\left(F(x)\right)\chi_1\left(a x\right).$$
Then the explicit values of $S_1$ are determined in the following.

\begin{theorem}\label{111}
Let $p$ be an odd prime. Let $S_1$ be defined above with $a \in V_n^{(p)}\setminus \{0\}$ and $F(x)$ be a vectorial dual-bent function satisfying Condition \uppercase\expandafter{\romannumeral1}. Then we have 
 \begin{eqnarray*}
 S_1=\left\{
\begin{array}{lll}
0  &   \mbox{if $F^*(a)=0$},\\
\varepsilon p^{\frac{n}{2}}    &   \mbox{if $\eta_m(F^*(a))=1$},\\
-\varepsilon p^{\frac{n}{2}}    &   \mbox{if $\eta_m(F^*(a))=-1$},\\
\end{array} \right.
 \end{eqnarray*}
 where $\varepsilon\in \{\pm 1\}$ was defined in Condition \uppercase\expandafter{\romannumeral1}.
\end{theorem}

\begin{IEEEproof}
By Condition \uppercase\expandafter{\romannumeral1}, $V_n^{(p)}=\gf_{p^{n_1}}\times \gf_{p^{n_2}}\times\cdots\times \gf_{p^{n_s}}$.
Let $a \in V_n^{(p)}\setminus \{0\}$. 
Using the Fourier expansion of the quadratic character $\eta_m$ by Lemma \ref{Fourier1}, we have 
 \begin{eqnarray}\label{eqn1}
 \nonumber 
S_1&=&\frac{1}{p^m}\sum_{x\in V_n^{(p)}}\left(\sum_{\lambda \in \widehat{\mathbb{F}}_{p^m}}G(\eta_m,\overline{\lambda})\lambda\left(F(x)\right)\right)\chi_1(ax)\\
 \nonumber
 &=&\frac{1}{p^m}\sum_{x\in V_n^{(p)}}\left(\sum_{y \in \gf_{p^m}}G(\eta_m,\overline{\lambda}_y)\lambda_1\left(yF(x)\right)\right)\chi_1(ax)\\
  \nonumber
 &=&\frac{1}{p^m}G(\eta_m,\overline{\lambda}_0)\sum_{x\in V_n^{(p)}}\chi_1(ax)+\frac{1}{p^m}\sum_{y\in \gf_{p^m}^*}G(\eta_m,\overline{\lambda}_y)\sum_{x\in V_n^{(p)}}\lambda_1(yF(x))\chi_1(ax)\\
 \nonumber
&=&0+\frac{1}{p^m}\sum_{y\in \gf_{p^m}^*}G(\eta_m,\overline{\lambda}_y)\sum_{x\in V_n^{(p)}}\lambda_1(yF(x))\chi_1(ax)\\
 \nonumber
&=&\frac{1}{p^m}G(\eta_m,\lambda_1)\sum_{y\in \gf_{p^m}^*}\eta_m(-y)\sum_{x \in V_n^{(p)}}\zeta_p^{\tr_{p^m/p}(yF(x))+\langle a,x \rangle_n}\\
 \nonumber
&=&\frac{1}{p^m}G(\eta_m,\lambda_1)\sum_{y\in \gf_{p^m}^*}\eta_m(-y)\sum_{x \in V_n^{(p)}}\zeta_p^{F_y(x)-\langle -a,x \rangle_n}\\
&=&\frac{1}{p^m}G(\eta_m,\lambda_1)\sum_{y\in \gf_{p^m}^*}\eta_m(-y)W_{F_y}(-a),
\end{eqnarray}
where $F_y(x):=\tr_{p^m/p}(yF(x))=\langle y, F(x)\rangle_m$  is the component function of $F$ and the firth equality holds due to $G(\eta_m,\overline{\lambda}_y)=\eta_m(-1)G(\eta_m,\lambda_y)=\eta_m(-y)G(\eta_m,\lambda_1)$ by Lemma \ref{quadGuasssum1}.
Since $p$ is an odd prime and $\gcd(d-1,p^m-1)=1$, it is easy to know that $d$ is even. By Condition \uppercase\expandafter{\romannumeral1}, $F_y$ is a weakly regular bent function.
By the definition of  weakly regular bent functions and Lemma \ref{lem8}, we have 
  \begin{eqnarray}\label{eqn2}
  W_{F_{y}}(-a)=\varepsilon p^{\frac{n}{2}}\zeta_{p}^{(F_y)^*(-a)}=\varepsilon p^{\frac{n}{2}}\zeta_{p}^{(F)^*_{y^{1-d}}(-a)}=\varepsilon p^{\frac{n}{2}}\zeta_{p}^{\tr_{p^m/p}(y^{1-d}F^*(-a))}=\varepsilon p^{\frac{n}{2}}\zeta_{p}^{\tr_{p^m/p}(y^{1-d}F^*(a))},
  \end{eqnarray}
where $(F)^*_{y^{1-d}}(-a)=\langle y^{1-d},F^*(-a)\rangle_{m}=\tr_{p^m/p}(y^{1-d}F^*(-a))$ and $F^*(-a)=(-1)^dF^*(a)=F^*(a)$.
Combining Equations (\ref{eqn1}) and (\ref{eqn2}), we have
 \begin{eqnarray*}
S_1&=&\frac{1}{p^m}\varepsilon p^{\frac{n}{2}}G(\eta_m,\lambda_1)\sum_{y\in \gf_{p^m}^*}\eta_m(-y)\zeta_p^{\tr_{p^m/p}(y^{1-d}F^*(a))}\\
&=&\frac{1}{p^m}\varepsilon p^{\frac{n}{2}}G(\eta_m,\lambda_1)\sum_{\widetilde{y}\in \gf_{p^m}^*}\eta_m(-\widetilde{y})\zeta_p^{\tr_{p^m/p}(\widetilde{y}F^*(a))}\\
&=&\frac{1}{p^m}\varepsilon p^{\frac{n}{2}}G(\eta_m,\lambda_1)\eta_m(-1)\sum_{\widetilde{y}\in \gf_{p^m}^*}\eta_m(\widetilde{y})\zeta_p^{\tr_{p^m/p}(\widetilde{y}F^*(a))},
 \end{eqnarray*}
 where $\widetilde{y}:=y^{1-d}$ also runs through $\gf_{p^m}^{*}$ if $y$ runs through $\gf_{p^m}^{*}$ as $\gcd(d-1,p^m-1)=1$ and $\eta_m(\widetilde{y})=\eta_m(y)$. 
 If $F^*(a)=0$, then 
\begin{eqnarray*}
S_1&=&\frac{1}{p^m}\varepsilon p^{\frac{n}{2}}G(\eta_m,\lambda_1)\eta_m(-1)\sum_{\widetilde{y}\in \gf_{p^m}^*}\eta_m(\widetilde{y})\\
&=&0
 \end{eqnarray*}
 by the orthogonal relation  of multiplicative characters.
 If $F^*(a)\neq0$, then 
 \begin{eqnarray*}
 S_1&=&\frac{1}{p^m}\varepsilon p^{\frac{n}{2}}G(\eta_m,\lambda_1)\eta_m(-1)G(\eta_m,\lambda_1)\eta_m(F^*(a))\\
 &=&\frac{1}{p^m}\varepsilon p^{\frac{n}{2}}G(\eta_m,\lambda_1)^{2}\eta_m(-1)\eta_m(F^*(a))\\
 &=&\left\{
\begin{array}{ll}
\frac{1}{p^m}\varepsilon p^{\frac{n}{2}}G(\eta_m,\lambda_1)^{2}\eta_m(-1)  &   \mbox{if $\eta_m(F^*(\alpha))=1$}\\
-\frac{1}{p^m}\varepsilon p^{\frac{n}{2}}G(\eta_m,\lambda_1)^{2}\eta_m(-1)    &   \mbox{if $\eta_m(F^*(\alpha))=-1$}\\
\end{array} \right.\\
&=&\left\{
\begin{array}{ll}
\varepsilon p^{\frac{n}{2}}  &   \mbox{if $\eta_m(F^*(a))=1$},\\
- \varepsilon p^{\frac{n}{2}}    &   \mbox{if $\eta_m(F^*(a))=-1$},\\
\end{array} \right.
\end{eqnarray*}
 where $G(\eta_m,\lambda_1)^2=\eta_m(-1)p^m$.
 Then the desired conclusion follows.
\end{IEEEproof}

In the following, we extend the hybrid character sums $S_1$ as follows. Let 
\begin{eqnarray*}
 \widehat{S}_1:=\sum_{x\in V_n^{(p)}}\psi(F(x))\chi_1(ax),
\end{eqnarray*}
where $\psi$ is a nontrivial multiplicative character of $\gf_{p^m}$ and $\chi_1$ is the canonical character of $V_n^{(p)}$ defined by $\chi_1(ax)=\zeta_{p}^{\langle a,x\rangle_n}$.

\begin{theorem}\label{222}
Let $p$ be an odd prime. Let $\widehat{S}_1$ be defined above with $a \in V_n^{(p)}\setminus \{0\}$ and $F(x)$ be a vectorial dual-bent function satisfying Condition \uppercase\expandafter{\romannumeral1}. If $\ord(\psi)>2$, then 
\begin{eqnarray*}
\widehat{S}_1=\left\{
\begin{array}{ll}
0  &   \mbox{if $F^*(a)=0$},\\
\frac{1}{p^m}\varepsilon p^{\frac{n}{2}}G(\psi,\lambda_1)\psi(-1)G(\overline{u'},\lambda_1)u'(F^{*}(a))    &   \mbox{if $F^*(a)\neq0$},\\
\end{array} \right.
\end{eqnarray*}
where $u'\in \widehat{\gf}_{p^m}^*$ satisfies $u'^{d-1}=\psi^{-1}$ for $\gcd(d-1,p^m-1)=1$ and $\varepsilon\in \{\pm 1\}$ was defined in Condition \uppercase\expandafter{\romannumeral1}. Besides, we have
\begin{eqnarray*}
|\widehat{S}_1|=\left\{
\begin{array}{ll}
0  &   \mbox{if $F^*(a)=0$},\\
p^{\frac{n}{2}}    &   \mbox{if $F^*(a)\neq0$}.\\
\end{array} \right.
\end{eqnarray*}
\end{theorem}

\begin{IEEEproof}
Using the Fourier expansion of the multiplicative character $\psi$ in Lemma \ref{Fourier1}, we have 
 \begin{eqnarray}\label{eqn3}
 \nonumber 
\widehat{S}_1&=&\frac{1}{p^m}\sum_{x\in V_n^{(p)}}\left(\sum_{\lambda \in \widehat{\mathbb{F}}_{p^m}}G(\psi,\overline{\lambda})\lambda\left(F(x)\right)\right)\chi_1(ax)\\
 \nonumber
 &=&\frac{1}{p^m}\sum_{x\in V_n^{(p)}}\left(\sum_{y \in \gf_{p^m}}G(\psi,\overline{\lambda}_y)\lambda_1\left(yF(x)\right)\right)\chi_1(ax)\\
  \nonumber
 &=&\frac{1}{p^m}G(\psi,\overline{\lambda}_0)\sum_{x\in V_n^{(p)}}\chi_1(ax)+\frac{1}{p^m}\sum_{y\in \gf_{p^m}^*}G(\psi,\overline{\lambda}_y)\sum_{x\in V_n^{(p)}}\lambda_1(yF(x))\chi_1(ax)\\
 \nonumber
&=&0+\frac{1}{p^m}\sum_{y\in \gf_{p^m}^*}G(\psi,\overline{\lambda}_y)\sum_{x\in V_n^{(p)}}\lambda_1(yF(x))\chi_1(ax)\\
 \nonumber
&=&\frac{1}{p^m}G(\psi,\lambda_1)\sum_{y\in \gf_{p^m}^*}\psi(-y^{-1})\sum_{x \in V_n^{(p)}}\zeta_p^{F_y(x)-\langle -a, x \rangle_n}\\
 \nonumber
&=&\frac{1}{p^m}G(\psi,\lambda_1)\sum_{y\in \gf_{p^m}^*}\psi(-y^{-1})W_{F_y}(-a),
\end{eqnarray}
where $G(\psi,\overline{\lambda}_y)=\psi(-1)\overline{\psi(y)}G(\psi,\lambda_1)=\psi(-y^{-1})G(\psi,\lambda_1)$ by Lemma \ref{quadGuasssum1} and $$F_y(x):=\tr_{p^m/p}(yF(x))=\langle y, F(x)\rangle_m$$  is the component function of $F$. 
According to Equation (\ref{eqn2}), we obtain 
\begin{eqnarray}\label{eq4}
 \nonumber 
\widehat{S}_1&=&\frac{1}{p^m}\varepsilon p^{\frac{n}{2}}G(\psi,\lambda_1)\sum_{y\in \gf_{p^m}^*}\psi(-y^{-1})\zeta_p^{\tr_{p^m/p}(y^{1-d}F^*(a))}\\
 \nonumber 
&=&\frac{1}{p^m}\varepsilon p^{\frac{n}{2}}G(\psi,\lambda_1)\sum_{y'\in \gf_{p^m}^*}\psi(-y')\zeta_p^{\tr_{p^m/p}(y'^{d-1}F^*(a))}\\
&=&\frac{1}{p^m}\varepsilon p^{\frac{n}{2}}G(\psi,\lambda_1)\sum_{y'\in \gf_{p^m}^*}\psi(-y')\lambda_1(y'^{d-1}F^*(a)),
\end{eqnarray}
where $y':=y^{-1}$. If $F^*(a)=0$, then 
\begin{eqnarray*}
\widehat{S}_1&=&\frac{1}{p^m}\varepsilon p^{\frac{n}{2}}G(\psi,\lambda_1)\sum_{y'\in \gf_{p^m}^*}\psi(-y')\\
&=&0
\end{eqnarray*}
as $\psi$ is nontrivial. 
If $F^*(a)\neq0$, by the Fourier expansion of $\lambda_1$ in Lemma \ref{Fourier2} and Equation (\ref{eq4}), we drive 
\begin{eqnarray*}
\widehat{S}_1&=&\frac{1}{p^m}\varepsilon p^{\frac{n}{2}}G(\psi,\lambda_1)\sum_{y'\in \gf_{p^m}^*}\psi(-y')\left(\frac{1}{p^m-1}\sum_{\mu\in \widehat{\gf}_{p^m}^{*}}G(\overline{\mu},\lambda_1)\mu(y'^{d-1}F^*(a))\right)\\
&=&\frac{1}{p^m(p^m-1)}\varepsilon p^{\frac{n}{2}}G(\psi,\lambda_1)\psi(-1)\sum_{\mu\in \widehat{\gf}_{p^m}^{*}}G(\overline{\mu},\lambda_1)\mu(F^*(a))\sum_{y'\in \gf_{p^m}^*}\psi(y')\mu(y'^{d-1})\\
&=&\frac{1}{p^m(p^m-1)}\varepsilon p^{\frac{n}{2}}G(\psi,\lambda_1)\psi(-1)\sum_{\mu\in \widehat{\gf}_{p^m}^{*}}G(\overline{\mu},\lambda_1)u(F^*(a))\sum_{y'\in \gf_{p^m}^*}\psi \mu^{d-1}(y')\\
&=&\frac{1}{p^m}\varepsilon p^{\frac{n}{2}}G(\psi,\lambda_1)\psi(-1)G(\overline{\mu'},\lambda_1)\mu'(F^*(a)),
\end{eqnarray*}
where the fourth equation holds due to the orthogonality relation of multiplicative characters and the fact that there exists an unique $\mu'\in\widehat{\mathbb{F}}_{p^m}^*$ such that $\mu'^{d-1}=\psi^{-1}$ as  $\gcd(d-1, p^m-1)=1$. By Lemma \ref{quadGuasssum2}, it is easy to know $\mid\widehat{S}_1\mid=p^{\frac{n}{2}}$ in this case.

Then the proof is completed.
\end{IEEEproof}

\subsection{The case that $F(x)$ satisfies Condition \uppercase\expandafter{\romannumeral2}}
In this subsection, we consider the case that $F(x)$ satisfies Condition \uppercase\expandafter{\romannumeral2} as follows.

\textbf{Condition \uppercase\expandafter{\romannumeral2}}: Let $p$ be an odd prime, $n, n_{j} (1\leq j\leq s)$, $m$, $t$ be positive integers such that $n=\sum_{j=1}^{s}n_{j}$, $t\mid n_{j}$, $t\mid m$, $2\mid (n-m)$, $n\geq 3m$ and $V_n^{(p)}=\gf_{p^{n_1}}\times \gf_{p^{n_2}}\times\cdots\times \gf_{p^{n_s}}$. Let $F: V_{n}^{(p)}\rightarrow \gf_{p^m}$ be a vectorial dual-bent function such that
 \begin{itemize}
\item{there is a vectorial dual $F^*$ satisfying ${F_c}^{*}=(F^*)_{c^{1-d}}$}, $c \in \gf_{p^m}^{*}$, where $\gcd(d-1,p^m-1)=1$;
\item{$F(ax)=a^{l}F(x)$, $a \in \gf_{p^t}^{*}$, $x \in V_{n}^{(p)}$ and $F(0)=0$, where $(l-1)(d-1)\equiv 1 \mod ({p^m-1})$};
\item{all component functions $F_c, c \in \gf_{p^m}^{*}$, are weakly regular with $\varepsilon_{F_{c}}=\upsilon\eta_m(c)$, $c \in \gf_{p^m}^{*}$, where $\upsilon \in \{\pm\epsilon^{m}\}$ is a constant for $\epsilon:=\sqrt{(-1)^{\frac{p-1}{2}}}$}.
\end{itemize}

The following lemma shows that the vectorial dual of a vectorial dual-bent function satisfying Condition \uppercase\expandafter{\romannumeral2} is also a vectorial dual-bent function satisfying Condition \uppercase\expandafter{\romannumeral2}.

\begin{lemma}\cite{WangJ2}\label{lem9}
Let $F$ be a vectorial dual-bent function satisfying Condition \uppercase\expandafter{\romannumeral2}. Then the vectorial dual $F^{*}$ with $(F_c)^{*}=(F^*)_{c^{1-d}}$, $c\in \gf_{p^m}^*$, is a vectorial dual-bent function such that $((F^*)_c)^*=F_{c^{1-l}}$, $c \in \gf_{p^m}$, $F^*(ax)=a^{d}F^*(x)$, $a \in \gf_{p^t}^*$, $F^*(0)=0$, and all component functions $(F^*)_c, c \in \gf_{p^m}^*$, are weakly regular with $\varepsilon_{(F^*)_c}=\upsilon^{-1}\eta_{m}(c).$
\end{lemma}

Now we recall some known classes
of vectorial dual-bent functions $F: V_{n}^{(p)}\rightarrow \gf_{p^m}$ satisfying Conditions \uppercase\expandafter{\romannumeral2} for general $m$. Note that when $\frac{n}{m}$ is odd, we deduce $\epsilon^n \in \{\pm\epsilon^m\}$.
\begin{itemize}
\item{Let $p$ be an odd prime and $t, m, s$ be positive integers with $t\mid m$. Let $s\geq3$ be odd. All non-degenerate quadratic forms $F: \gf_{p^m}^s\rightarrow \gf_{p^m}$ or $ F: \gf_{p^{ms}}\rightarrow \gf_{p^m}$ are vectorial dual-bent functions satisfying Condition \uppercase\expandafter{\romannumeral2} with $l=d=2$ \cite{WangJ}, \cite{WangJ1}. The followings are two families of such functions.
    \begin{enumerate}
    \item{Let $m, n, t$ be positive integers such that  $t\mid m, m\mid n, \frac{n}{m}\geq3$ is odd, $a \in \gf_{p^n}^*$. Define $F: \gf_{p^n}\rightarrow \gf_{p^m}$ by
 \begin{eqnarray}\label{Tr(x^2)1}
 F(x)=\tr_{p^n/p^m}(a x^2).
\end{eqnarray}
Then $F$ is a vectorial dual-bent function satisfying Condition \uppercase\expandafter{\romannumeral2} with $$l=d=2,\upsilon=(-1)^{n-1}\epsilon^n\eta_n(a).$$
        }
     \item{Let $m, t, s$ be positive integers such that $t\mid m, s\geq3$ is odd, $\alpha_i \in \gf_{p^m}^*, 1\leq i \leq s$. Define $F: \gf_{p^m}^s\rightarrow \gf_{p^m}$ by
      \begin{eqnarray*}
      F(x_1,\ldots, x_s)=\sum_{i=1}^{s}\alpha_ix_i^2.
      \end{eqnarray*} 
      Then $F$ is a vectorial dual-bent function satisfying Condition   \uppercase\expandafter{\romannumeral2} with $$l=d=2,\upsilon=(-1)^{m-1}\epsilon^{ms}\eta_m(\alpha_1\cdots\alpha_s).$$
     }       
    \end{enumerate}
    }
    \item{Let $p$ be an odd prime, $m, t, n', n''$ be positive integers such that $t\mid m, m\mid n', m\mid n'', \frac{n'}{m}$ is odd. For $i \in \gf_{p^m}$, let $H(i,x): \gf_{p^{n'}}\rightarrow \gf_{p^m}$ be defined by $H(0,x)=\tr_{p^{n'}/p^m}(\alpha_1x^2), H(i,x)=\tr_{p^{n'}/p^m}(\alpha_2x^2)$ if $i$ is a square in $\gf_{p^m}^*$, $H(i,x)=\tr_{p^{n'}/p^m}(\alpha_3x^2)$ if $i$ is a non-square in $\gf_{p^m}^*$, where $\alpha_1, \alpha_2, \alpha_3$ are all square elements  or all non-square elements in $\gf_{p^{n'}}^*$. Let $G: \gf_{p^{n''}}\times\gf_{p^{n''}}\rightarrow \gf_{p^m}$ be defined by $G(y_1,y_2)= \tr_{p^{n''}/p^m}(\beta y_1L(y_2))$, where $\beta \in \gf_{p^{n''}}^*$, $L(x)=\sum a_ix^{p^{mi}}$ is a $p^m$- polynomial over $\gf_{p^{n''}}$ inducing a permutation of $\gf_{p^{n''}}$. Let $\gamma \in \gf_{p^{n''}}^*$ and $F: \gf_{p^{n'}}\times \gf_{p^{n''}}\times \gf_{p^{n''}}\rightarrow \gf_{p^m}$ be defined by
    \begin{eqnarray*}
    F(x,y_1,y_2)=H(\tr_{p^{n''}/p^m}(\gamma y_2^2),x)+G(y_1,y_2).
    \end{eqnarray*}
    Then $F$ is a vectorial dual-bent function satisfying Condition\uppercase\expandafter{\romannumeral2} with $l=d=2, \upsilon=(-1)^{n'-1}\epsilon^{n'}\eta_{n'}(\alpha_1)$ \cite{WangJ1}.
    }
\end{itemize}

In the following, let $F(x)$ be a vectorial dual-bent function satisfying Condition \uppercase\expandafter{\romannumeral2} and $ \in V_n^{(p)}\setminus \{0\}$. Denote by
 \begin{eqnarray*}
 S_2=\sum_{x\in V_n^{(p)}}\eta_{m}(F(x))\chi_1(ax).
 \end{eqnarray*}
 The explict value of $S_2$ is determiend in the following theorem.

\begin{theorem}\label{333}
Let $p$ be an odd prime. Let $S_2$ be defined above for $a \in V_n^{(p)}\setminus \{0\}$ and $F(x)$ be a vectorial dual-bent function satisfying Condition \uppercase\expandafter{\romannumeral2}. Then we have 
\begin{eqnarray*}
 S_2=\left\{
\begin{array}{ll}
\frac{p^m-1}{p^m}\upsilon p^{\frac{n}{2}} (-1)^{m-1}\epsilon^m\sqrt{p^m}\eta_m(-1)   &   \mbox{if $F^*(a)=0$},\\
-\frac{1}{p^m}\upsilon p^{\frac{n}{2}} (-1)^{m-1}\epsilon^m\sqrt{p^m}\eta_m(-1)    &   \mbox{if $F^*(a)\neq0$},
\end{array} \right.
\end{eqnarray*}
where $\upsilon \in \{\pm\epsilon^{m}\}$ is a constant defined in Condition \uppercase\expandafter{\romannumeral2} for $\epsilon:=\sqrt{(-1)^{\frac{p-1}{2}}}$.
\end{theorem}

\begin{IEEEproof}
Using the Fourier expansion of the quadratic character $\eta_m$ in Lemma \ref{Fourier1}, we have 
 \begin{eqnarray}\label{eqn5}
 \nonumber 
S_2&=&\frac{1}{p^m}\sum_{x\in V_n^{(p)}}\left(\sum_{\lambda \in \widehat{\mathbb{F}}_{p^m}}G(\eta_m,\overline{\lambda})\lambda\left(F(x)\right)\right)\chi_1(ax)\\
 \nonumber
 &=&\frac{1}{p^m}\sum_{x\in V_n^{(p)}}\left(\sum_{y \in \gf_{p^m}}G(\eta_m,\overline{\lambda}_y)\lambda\left(yF(x)\right)\right)\chi_1(ax)\\
  \nonumber
 &=&\frac{1}{p^m}G(\eta_m,\overline{\lambda}_0)\sum_{x\in V_n^{(p)}}\chi_\alpha(x)+\frac{1}{p^m}\sum_{y\in \gf_{p^m}^*}G(\eta_m,\overline{\lambda}_y)\sum_{x\in V_n^{(p)}}\lambda_1(yF(x))\chi_1(ax)\\
 \nonumber
&=&0+\frac{1}{p^m}\sum_{y\in \gf_{p^m}^*}G(\eta_m,\overline{\lambda}_y)\sum_{x\in V_n^{(p)}}\lambda_1(yF(x))\chi_1(ax)\\
 \nonumber
&=&\frac{1}{p^m}G(\eta_m,\lambda_1)\sum_{y\in \gf_{p^m}^*}\eta_m(-y)\sum_{x \in V_n^{(p)}}\zeta_p^{\tr_{p^m/p}(yF(x))-\langle a, x \rangle_n}\\
 \nonumber
&=&\frac{1}{p^m}G(\eta_m,\lambda_1)\sum_{y\in \gf_{p^m}^*}\eta_m(-y)\sum_{x \in V_n^{(p)}}\zeta_p^{F_y(x)-\langle a, x \rangle_n}\\
&=&\frac{1}{p^m}G(\eta_m,\lambda_1)\sum_{y\in \gf_{p^m}^*}\eta_m(-y)W_{F_y}(-a),
\end{eqnarray}
where $F_y(x):=\tr_{p^m/p}(yF(x))=\langle y, F(x)\rangle_m$  is the component function of $F$ and the firth equality holds due to $G(\eta_m,\overline{\lambda}_y)=\eta_m(-1)G(\eta_m,\lambda_y)=\eta_m(-y)G(\eta_m,\lambda_1)$ by Lemma \ref{quadGuasssum1}.
Since $p$ is an odd prime, it is easy to know that $d$ is even by Condition \uppercase\expandafter{\romannumeral2}. By Condition \uppercase\expandafter{\romannumeral2}, 
$F_y, y \in \gf_{p^m}^{*}$, are weakly regular with $\varepsilon_{F_{y}}=\upsilon\eta_m(y)$.
According to the definition of the weakly regular bent function and Lemma \ref{lem9}, we have 
  \begin{eqnarray}\label{eqn6}
  W_{F_{y}}(-a)=\upsilon\eta_m(y) p^{\frac{n}{2}}\zeta_{p}^{(F)^*_{y^{1-d}}(-a)}=\upsilon\eta_m(y) p^{\frac{n}{2}}\zeta_{p}^{\tr_{p^m/p}(y^{1-d}F^*(-a))}=\upsilon\eta_m(y) p^{\frac{n}{2}}\zeta_{p}^{\tr_{p^m/p}(y^{1-d}F^*(a))},
  \end{eqnarray}
where $(F)^*_{y^{1-d}}(-a)=\langle y^{1-d},F^*(-a)\rangle_{m}=\tr_{p^m/p}(y^{1-d}F^*(-a))$ and $F^*(-a)=(-1)^dF^*(a)=F^*(a)$.
Combining Equations (\ref{eqn5}) and (\ref{eqn6}), we have
   \begin{eqnarray*}
S_2&=&\frac{1}{p^m}\upsilon p^{\frac{n}{2}}G(\eta_m,\lambda_1)\sum_{y\in \gf_{p^m}^*}\eta_m(-y)\eta_m(y)\zeta_p^{\tr_{p^m/p}(y^{1-d}F^*(a))}\\
&=&\frac{1}{p^m}\upsilon p^{\frac{n}{2}}G(\eta_m,\lambda_1)\eta_m(-1)\sum_{z\in \gf_{p^m}^*}\zeta_p^{\tr_{p^m/p}(zF^*(a))},
 \end{eqnarray*}
 where $z:=y^{1-d}$ runs through $\gf_{p^m}^*$ if $y$ runs through $\gf_{p^m}^*$ due to $\gcd(d-1,p^m-1)=1$. Then we obtain 
\begin{eqnarray*}
 S_2=\left\{
\begin{array}{ll}
\frac{p^m-1}{p^m}\upsilon p^{\frac{n}{2}} G(\eta_m,\lambda_1)\eta_m(-1)   &   \mbox{if $F^*(a)=0$},\\
-\frac{1}{p^m}\upsilon p^{\frac{n}{2}} G(\eta_m,\lambda_1)\eta_m(-1)    &   \mbox{if $F^*(a)\neq0$},
\end{array} \right.
\end{eqnarray*}
by the orthogonal relation of additive characters. 
By Lemma \ref{quadGuasssum3}, we have
$$G(\eta_m,\lambda_1)=(-1)^{m-1}(\sqrt{-1})^{(\frac{p-1}{2})^2m}\sqrt{p^m}=(-1)^{m-1}\epsilon^m\sqrt{p^m}.$$ Thus 
\begin{eqnarray*}
 S_2=\left\{
\begin{array}{ll}
\frac{p^m-1}{p^m}\upsilon p^{\frac{n}{2}} (-1)^{m-1}\epsilon^m\sqrt{p^m}\eta_m(-1)   &   \mbox{if $F^*(a)=0$},\\
-\frac{1}{p^m}\upsilon p^{\frac{n}{2}} (-1)^{m-1}\epsilon^m\sqrt{p^m}\eta_m(-1)    &   \mbox{if $F^*(a)\neq0$}.
\end{array} \right.
\end{eqnarray*}
Then the desired conclusion follows.
\end{IEEEproof}

In the following, we study the hybrid sums given as
\begin{eqnarray*}
 \widehat{S}_2:=\sum_{x\in V_n^{(p)}}\psi(F(x))\chi_1(ax),
\end{eqnarray*}
where  $a \in V_n^{(p)}\setminus \{0\}$, $\psi$ is a nontrivial multiplicative of $\gf_{p^m}$ and $\chi_1$ is the canonical additive character of $V_n^{(p)}$.

\begin{theorem}
Let $p$ be an odd prime. Let $\widehat{S}_2$ be defined above with $a \in V_n^{(p)}\setminus \{0\}$ and $F(x)$ be a vectorial dual-bent function satisfying Condition \uppercase\expandafter{\romannumeral2}. If $\ord(\psi)>2$, then we have
\begin{eqnarray*}
\widehat{S}_2=\left\{
\begin{array}{ll}
0  &   \mbox{if $F^*(a)=0$},\\
\frac{1}{p^m}\upsilon p^{\frac{n}{2}}G(\psi,\lambda_1)\psi(-1)G(\overline{\mu'},\lambda_1)\mu'(F^{*}(a))    &   \mbox{if $F^*(a)\neq0$},\\
\end{array} \right.
\end{eqnarray*}
where $\mu'\in \widehat{\gf}_{p^m}^*$ such that $\mu'^{d-1}=\psi^{-1}$ for $\gcd(d-1, p^m-1)=1$ and $\upsilon \in \{\pm\epsilon^{m}\}$ is a constant defined in Condition \uppercase\expandafter{\romannumeral2} for $\epsilon:=\sqrt{(-1)^{\frac{p-1}{2}}}$. Besides, we have
\begin{eqnarray*}
\mid\widehat{S}_2\mid=\left\{
\begin{array}{ll}
0  &   \mbox{if $F^*(a)=0$},\\
p^{\frac{n}{2}}    &   \mbox{if $F^*(a)\neq0$}.\\
\end{array} \right.
\end{eqnarray*}
\end{theorem}
\begin{IEEEproof}
Using the Fourier expansion of the multiplicative character $\psi$ by Lemma \ref{Fourier1}, we have 
 \begin{eqnarray*}\label{eqn7}
 \nonumber 
\widehat{S}_2&=&\frac{1}{p^m}\sum_{x\in V_n^{(p)}}\left(\sum_{\lambda \in \widehat{\gf}_{p^m}}G(\psi,\overline{\lambda})\lambda\left(F(x)\right)\right)\chi_1(ax)\\
 \nonumber
 &=&\frac{1}{p^m}\sum_{x\in V_n^{(p)}}\left(\sum_{y \in \gf_{p^m}}G(\psi,\overline{\lambda}_y)\lambda\left(yF(x)\right)\right)\chi_1(ax)\\
  \nonumber
 &=&\frac{1}{p^m}G(\psi,\overline{\lambda}_0)\sum_{x\in V_n^{(p)}}\chi_1(ax)+\frac{1}{p^m}\sum_{y\in \gf_{p^m}^*}G(\psi,\overline{\lambda}_y)\sum_{x\in V_n^{(p)}}\lambda_1(yF(x))\chi_1(ax)\\
 \nonumber
&=&0+\frac{1}{p^m}\sum_{y\in \gf_{p^m}^*}G(\psi,\overline{\lambda}_y)\sum_{x\in V_n^{(p)}}\lambda_1(yF(x))\chi_1(ax)\\
 \nonumber
&=&\frac{1}{p^m}G(\psi,\lambda_1)\sum_{y\in \gf_{p^m}^*}\psi(-y^{-1})\sum_{x \in V_n^{(p)}}\zeta_p^{\tr_{p^m/p}(yF(x))-\langle -a, x \rangle_n}\\
 \nonumber
&=&\frac{1}{p^m}G(\psi,\lambda_1)\sum_{y\in \gf_{p^m}^*}\psi(-y^{-1})\sum_{x \in V_n^{(p)}}\zeta_p^{F_y(x)-\langle -a, x \rangle_n}\\
&=&\frac{1}{p^m}G(\psi,\lambda_1)\sum_{y\in \gf_{p^m}^*}\psi(-y^{-1})W_{F_y}(-a),
\end{eqnarray*}
where $G(\psi,\overline{\lambda}_y)=\psi(-1)\overline{\psi(y)}G(\psi,\lambda_1)=\psi(-y^{-1})G(\psi,\lambda_1)$ by Lemma \ref{quadGuasssum1} and $$F_y(x):=\tr_{p^m/p}(yF(x))=\langle y, F(x)\rangle_m$$  is the component function of $F$. By Equation (\ref{eqn6}), we then obtain 
\begin{eqnarray*}
\widehat{S}_2&=&\frac{1}{p^m}\upsilon p^{\frac{n}{2}}G(\psi,\lambda_1)\sum_{y\in \gf_{p^m}^*}\psi(-y^{-1})\eta_m(y)\zeta_p^{\tr_{p^m/p}(y^{1-d}F^*(a))}\\
&=&\frac{1}{p^m}\upsilon p^{\frac{n}{2}}G(\psi,\lambda_1)\psi(-1)\sum_{y\in \gf_{p^m}^*}\psi(y^{-1})\eta_m(y^{-1})\zeta_p^{\tr_{p^m/p}(y^{1-d}F^*(a))}\\
&=&\frac{1}{p^m}\upsilon p^{\frac{n}{2}}G(\psi,\lambda_1)\psi(-1)\sum_{y'\in \gf_{p^m}^*}\psi\eta_m(y')\zeta_p^{\tr_{p^m/p}((y')^{d-1}F^*(a)})\\
&=&\frac{1}{p^m}\upsilon p^{\frac{n}{2}}G(\psi,\lambda_1)\psi(-1)\sum_{y'\in \gf_{p^m}^*}\psi\eta_m(y')\lambda_1((y')^{d-1}F^*(a)),
 \end{eqnarray*}
 where $y':=y^{-1}$. Note that $\psi\neq\eta_m$ as $\ord(\psi)>2$.
If $F^*(a)=0$, then $\widehat{S}_2=0$ as $\psi\neq\eta_m$.
If $F^*(a)\neq0$, by the Fourier expansion of $\lambda_1$ in Lemma \ref{Fourier2}, then we drive 
\begin{eqnarray*}
\widehat{S}_2&=&\frac{1}{p^m}\upsilon p^{\frac{n}{2}}G(\psi,\lambda_1)\psi(-1)\sum_{y'\in \gf_{p^m}^*}\psi\eta_m(y')\left(\frac{1}{p^m-1}\sum_{\mu\in \widehat{\gf}_{p^m}^{*}}G(\overline{\mu},\lambda_1)\mu((y')^{d-1}F^*(a))\right)\\
&=&\frac{1}{p^m(p^m-1)}\upsilon p^{\frac{n}{2}}G(\psi,\lambda_1)\psi(-1)\sum_{\mu\in \widehat{\gf}_{p^m}^{*}}G(\overline{\mu},\lambda_1)\mu(F^*(a))\sum_{y'\in \gf_{p^m}^*}\psi\eta_m(y')\mu((y')^{d-1})\\
&=&\frac{1}{p^m(p^m-1)}\upsilon p^{\frac{n}{2}}G(\psi,\lambda_1)\psi(-1)\sum_{\mu\in \widehat{\gf}_{p^m}^{*}}G(\overline{\mu},\lambda_1)\mu(F^*(a))\sum_{y'\in \gf_{p^m}^*}\psi \eta_m\mu^{d-1}(y')\\
&=&\frac{1}{p^m}\upsilon p^{\frac{n}{2}}G(\psi,\lambda_1)\psi(-1)G(\overline{\mu'},\lambda_1)\mu'(F^*(a)),
\end{eqnarray*}
where the fourth equation holds due to the orthogonality relation of multiplicative characters and the fact that
there existis an unique $\mu' \in \widehat{\gf}_{p^m}^*$ satisfying $\mu'^{d-1}=\psi^{-1}\eta_m$  for $\gcd(d-1, p^m-1)=1$. By Lemma \ref{quadGuasssum2}, it is easy to know $\mid\widehat{S}_2\mid=p^{\frac{n}{2}}$.
Then the desired conclusion follows.
\end{IEEEproof}

\section{Application of the hybrid character sums in codebooks}
In this section, we shall construct three families of asymptotically optimal codebooks with small alphabet sizes. 
The hybrid chararacter sums studied in this paper play an important role in determining the maximal cross-correlation
amplitudes of these families of codebooks. 
\subsection{Some lemmas}
In this subsection, we present some lemmas which will be used to prove the main results of this section. 
\begin{lemma}\label{lem14}
Let $T:=\sum\limits_{\substack{x \in V_n^{(p)}\\F(x)=0}}\chi_1(ax)$, $a \in V_{n}^{(p)}\setminus\{0\}$, $p$ be an odd prime and $F(x)$ be a vectorial dual-bent function satisfying Condition \uppercase\expandafter{\romannumeral1}. Then we have
\begin{eqnarray*}
T=\left\{
\begin{array}{ll}
\frac{p^m-1}{p^m}\varepsilon p^{\frac{n}{2}}  &   \mbox{if $F^*(a)=0$},\\
-\frac{\varepsilon p^{\frac{n}{2}}}{p^m}   &   \mbox{if $F^*(a)\neq0$},\\
\end{array} \right.
\end{eqnarray*}
 where $\varepsilon\in \{\pm 1\}$ was defined in Condition \uppercase\expandafter{\romannumeral1}.
\end{lemma}

\begin{IEEEproof}
By the orthogonal relation of additive characters, we have
\begin{eqnarray*}
T&=&\frac{1}{p^m}\sum_{x \in V_n^{(p)}}\chi_1(ax)\sum_{y \in \gf_{p^m}}\lambda_{1}(yF(x))\\
&=&0+\frac{1}{p^m}\sum_{y \in \gf_{p^m}^{*}}\sum_{x \in V_n^{(p)}}\zeta_{p}^{\tr_{p^m/p}(yF(x))-\langle -a, x \rangle_n}\\
&=&\frac{1}{p^m}\sum_{y \in \gf_{p^m}^{*}}\sum_{x \in V_n^{(p)}}\zeta_{p}^{F_y(x)-\langle -a, x \rangle_n}\\
&=&\frac{1}{p^m}\sum_{y \in \gf_{p^m}^*}W_{F_y}(-a),
\end{eqnarray*}
where $\lambda_1$ is the canonical additive character of $\gf_{p^m}$ and $F_y(x):=\tr_{p^m/p}(yF(x))=\langle y, F(x)\rangle_m$  is the component function of $F$.
By Equation (\ref{eqn2}), we have 
\begin{eqnarray*}
T&=&\frac{1}{p^m}\varepsilon p^{\frac{n}{2}}\sum_{y \in \gf_{p^m}^*}\zeta_{p}^{\tr_{p^m/p}(y^{1-d}F^*(a))}\\
&=&\frac{1}{p^m}\varepsilon p^{\frac{n}{2}}\sum_{\widetilde{z} \in \gf_{p^m}^*}\zeta_{p}^{\tr_{p^m/p}(\widetilde{z}F^*(a))}\\
&=&\left\{
\begin{array}{ll}
\frac{p^m-1}{p^m}\varepsilon p^{\frac{n}{2}}  &   \mbox{if $F^*(a)=0$},\\
-\frac{\varepsilon p^{\frac{n}{2}}}{p^m}   &   \mbox{if $F^*(a)\neq0$}.\\
\end{array} \right.
\end{eqnarray*}
where $\widetilde{z}:=y^{1-d}$ and $\widetilde{z}$ runs through $\gf_{p^m}$ if $y$ runs through $\gf_{p^m}$ as $\gcd(d-1,p^m-1)=1$. The proof is completed.
\end{IEEEproof}

\begin{lemma}\label{lem15}
Let $T':=\sum\limits_{\substack{x \in V_n^{(p)}\\F(x)=0}}\chi_1(ax)$, $a \in V_{n}^{(p)}\setminus\{0\}$, $p$ be an odd prime and $F(x)$ be a vectorial dual-bent function satisfying Condition \uppercase\expandafter{\romannumeral2}. Then 
\begin{eqnarray*}
T'&=&\left\{
\begin{array}{lll}
0  &   \mbox{if $F^*(a)=0$},\\
\frac{1}{p^m}\upsilon p^{\frac{n}{2}}(-1)^{m-1}\epsilon^m\sqrt{p^m}  &   \mbox{if $\eta_m(F^*(a))=1$},\\
-\frac{1}{p^m}\upsilon p^{\frac{n}{2}}(-1)^{m-1}\epsilon^m\sqrt{p^m}   &   \mbox{if $\eta_m(F^*(a))=-1$},\\
\end{array} \right.
\end{eqnarray*}
where $\upsilon \in \{\pm\epsilon^{m}\}$ is a constant defined in Condition \uppercase\expandafter{\romannumeral2} for $\epsilon:=\sqrt{(-1)^{\frac{p-1}{2}}}$.
\end{lemma}

\begin{IEEEproof}
By the orthogonal relation of additive characters, we have
\begin{eqnarray*}
T'&=&\frac{1}{p^m}\sum_{x \in V_n^{(p)}}\chi_a(x)\sum_{y \in \gf_{p^m}}\lambda_{1}(yF(x))\\
&=&0+\frac{1}{p^m}\sum_{y \in \gf_{p^m}^{*}}\sum_{x \in V_n^{(p)}}\zeta_{p}^{yF(x)-\langle -a, x \rangle_n}\\
&=&\frac{1}{p^m}\sum_{y \in \gf_{p^m}^*}W_{F_y}(-a),
\end{eqnarray*}
where $F_y(x):=\tr_{p^m/p}(yF(x))=\langle y, F(x)\rangle_m$  is the component function of $F$.
By Equation (\ref{eqn6}), we then have 
\begin{eqnarray*}
T'&=&\frac{1}{p^m}\upsilon p^{\frac{n}{2}}\sum_{y \in \gf_{p^m}^*}\eta_{m}(y)\zeta_{p}^{\tr_{p^m/p}(y^{1-d}F^*(a))}\\
&=&\frac{1}{p^m}\upsilon p^{\frac{n}{2}}\sum_{\widetilde{y} \in \gf_{p^m}^*}\eta_{m}(\widetilde{y})\zeta_{p}^{\tr_{p^m/p}(\widetilde{y}F^*(a))}\\
&=&\frac{1}{p^m}\upsilon p^{\frac{n}{2}}\sum_{\widetilde{y} \in \gf_{p^m}^*}\eta_{m}(\widetilde{y})\lambda_1(\widetilde{y}F^*(a)),
\end{eqnarray*}
where $\widetilde{y}:=y^{1-d}$, $\widetilde{y}$ runs through $\gf_{p^m}^*$ if $y$ runs through $\gf_{p^m}^*$ and $\eta_m(\widetilde{y})=\eta_m(y)$ due to $\gcd(d-1,p^m-1)=1$.
By the orthogonal relation of multiplicative characters and the definition of Gaussian sums, we further derive 
\begin{eqnarray*}
T'&=&\left\{
\begin{array}{ll}
0  &   \mbox{if $F^*(a)=0$}\\
\frac{1}{p^m}\upsilon p^{\frac{n}{2}}G(\eta_m,\lambda_1)\eta_m(F^*(a))  &   \mbox{if $F^*(a)\neq0$}\\
\end{array} \right.\\
&=&\left\{
\begin{array}{lll}
0  &   \mbox{if $F^*(a)=0$},\\
\frac{1}{p^m}\upsilon p^{\frac{n}{2}}G(\eta_m,\lambda_1)  &   \mbox{if $\eta_m(F^*(a))=1$},\\
-\frac{1}{p^m}\upsilon p^{\frac{n}{2}}G(\eta_m,\lambda_1)   &   \mbox{if $\eta_m(F^*(a))=-1$}.\\
\end{array} \right.
\end{eqnarray*}
By Lemma \ref{quadGuasssum3}, we know $G(\eta_m,\lambda_1)=(-1)^{m-1}(\sqrt{-1})^{(\frac{p-1}{2})^2m}\sqrt{p^m}=(-1)^{m-1}\epsilon^m\sqrt{p^m}$. Thus 
\begin{eqnarray*}
T'&=&\left\{
\begin{array}{lll}
0  &   \mbox{if $F^*(\alpha)=0$},\\
\frac{1}{p^m}\upsilon p^{\frac{n}{2}}(-1)^{m-1}\epsilon^m\sqrt{p^m}  &   \mbox{if $\eta_m(F^*(\alpha))=1$},\\
-\frac{1}{p^m}\upsilon p^{\frac{n}{2}}(-1)^{m-1}\epsilon^m\sqrt{p^m}   &   \mbox{if $\eta_m(F^*(\alpha))=-1$}.\\
\end{array} \right.
\end{eqnarray*}
The proof is completed.
\end{IEEEproof}

Define 
\begin{eqnarray*}
N_0(F)&=&|x \in V_{n}^{(p)}: F(x)=0|,\\
N_1(F)&=&| x \in V_{n}^{(p)}: \eta_m(F(x))=1|,\\
N_{-1}(F)&=&| x \in V_{n}^{(p)}: \eta_m(F(x))=-1|,
\end{eqnarray*}
where $F(x)$ is  a function from $V_{n}^{(p)}$ to $\gf_{p^m}$. Then we have the following results on $N_i(F)$ for $i=0,\pm 1$.

\begin{lemma}\cite{WangJ2}\label{lem16}
If $F(x)$ is a vectorial dual-bent function satisfying Condition \uppercase\expandafter{\romannumeral1}, then 
\begin{eqnarray*}
N_0(F)=p^{n-m}+\varepsilon p^{\frac{n}{2}-m}(p^m-1),
\end{eqnarray*}
 where $\varepsilon\in \{\pm 1\}$ was defined in Condition \uppercase\expandafter{\romannumeral1}.
If $F(x)$ is a vectorial dual-bent function satisfying Condition \uppercase\expandafter{\romannumeral2}, then 
\begin{eqnarray*}
N_0(F)=p^{n-m}.
\end{eqnarray*}
\end{lemma}

\begin{lemma}\label{lem17}
Let $F(x)$ be a vectorial dual-bent function satisfying Condition \uppercase\expandafter{\romannumeral1}, then 
\begin{eqnarray*}
N_{1}(F)=N_{-1}(F)=\frac{(p^n-\varepsilon p^{\frac{n}{2}})(p^m-1)}{2p^m},
\end{eqnarray*}
where $\varepsilon\in \{\pm 1\}$ was defined in Condition \uppercase\expandafter{\romannumeral1}.
\end{lemma}

\begin{IEEEproof}
It is easy to know that 
\begin{eqnarray}\label{eq7}
N_0(F)+N_1(F)+N_{-1}(F)=p^n.
\end{eqnarray}
Denote by 
\begin{eqnarray}\label{eq8}
\nonumber \Omega &=&\sum_{x \in V_{n}^{(p)}}\left(\sum_{y \in \gf_{p^m}}\lambda_1(y^2F(x))\right)\\
\nonumber &=&N_0(F)p^m+\sum_{\substack{x \in V_{n}^{(p)}\\F(x)\neq 0}}\eta_m(F(x))\left(\sum_{y \in \gf_{p^m}}\lambda_1(y^2F(x))\eta_m(y^2F(x))\right)\\
&=&N_0(F)p^m+(N_1(F)-N_{-1}(F))G(\eta_m,\lambda_1),
\end{eqnarray}
where $\eta_m(0)=0$.
In addition, we also have
\begin{eqnarray*}
\Omega&=&p^n+\sum_{y \in \gf_{p^m}^{*}}\sum_{x \in V_n^{(p)}}\zeta_p^{\tr_{p^m/p}(y^2F(x))}\\
&=&p^n+\sum_{y \in \gf_{p^m}^{*}}W_{F_{y^2}}(0)\\
&=&p^n+\varepsilon p^{\frac{n}{2}}(p^m-1),
\end{eqnarray*}
where $F_{y^2}(x)$ is weakly regular bent and $W_{F_{y^2}}(0)=\varepsilon p^{\frac{n}{2}}$ due to the definition of weakly bent function and $F*(0)=0$ by Lemma \ref{lem8}.
Combining Equation (\ref{eq7}), Equation (\ref{eq8}) and Lemma \ref{lem16}, we have
\begin{eqnarray*}
\left\{
\begin{array}{ll}
N_0(F)+N_1(F)+N_{-1}(F)=p^n,  \\
 N_0(F)p^m+(N_1(F)-N_{-1}(F))G(\eta_m,\lambda_1)=p^n+\varepsilon p^{\frac{n}{2}}(p^m-1).
\end{array} \right.\\
\end{eqnarray*}
Solving the above system of equations, we derive
\begin{eqnarray*}
N_{1}(F)=N_{-1}(F)=\frac{(p^n-\varepsilon p^{\frac{n}{2}})(p^m-1)}{2p^m}.
\end{eqnarray*}
The proof is completed.
\end{IEEEproof}

\begin{lemma}\label{lem19}
Let $F(x)$ be a vectorial dual-bent function satisfying Condition \uppercase\expandafter{\romannumeral1} and $F^*(x)$ be its dual. Then 
\begin{eqnarray*}
N_{0}(F^*)&=&p^{n-m}+\varepsilon p^{\frac{n}{2}-m}(p^m-1), \\
N_{1}(F^*)&=&N_{-1}(F^*)=\frac{(p^n-\varepsilon p^{\frac{n}{2}})(p^m-1)}{2p^m},
\end{eqnarray*}
where $\varepsilon\in \{\pm 1\}$ was defined in Condition \uppercase\expandafter{\romannumeral1}.
\end{lemma}
\begin{IEEEproof}
$N_{0}(F^*)$ was determined in \cite{WangJ2}.  For $N_{1}(F^*)$ and $N_{-1}(F^*)$, they can be determined in a similar way as that of Lemma \ref{lem17}. The proof is omitted here. 
\end{IEEEproof}

\begin{lemma}
Let $F(x)$ be a vectorial dual-bent function satisfying Condition \uppercase\expandafter{\romannumeral2}. Then 
\begin{eqnarray*}
N_{1}(F)=\frac{(p^{n-m}+(-1)^{m-1}\epsilon^m\eta_m(-1)\upsilon p^{\frac{n-m}{2}})(p^m-1)}{2},\\
N_{-1}(F)=\frac{(p^{n-m}-(-1)^{m-1}\epsilon^m\eta_m(-1)\upsilon p^{\frac{n-m}{2}})(p^m-1)}{2},
\end{eqnarray*}
where $\upsilon \in \{\pm\epsilon^{m}\}$ is a constant defined in Condition \uppercase\expandafter{\romannumeral2} for $\epsilon:=\sqrt{(-1)^{\frac{p-1}{2}}}$. 
\end{lemma}
\begin{IEEEproof}
The proof is similar to that of Lemma \ref{lem17} and omitted here.
\end{IEEEproof}

\begin{lemma}\label{lem20}
Let $F(x)$ be a vectorial dual-bent function with Condition \uppercase\expandafter{\romannumeral2} and $F^*(x)$ be its dual. Then 
\begin{eqnarray*}
N_{0}(F^*)&=&p^{n-m},\\
N_{1}(F^*)&=&\frac{(p^{n-m}+(-1)^{m-1}\epsilon^m\upsilon^{-1}\eta_m(-1) p^{\frac{n-m}{2}})(p^m-1)}{2},\\
N_{-1}(F^*)&=&\frac{(p^{n-m}-(-1)^{m-1}\epsilon^m\upsilon^{-1}\eta_m(-1) p^{\frac{n-m}{2}})(p^m-1)}{2},
\end{eqnarray*}
where $\upsilon \in \{\pm\epsilon^{m}\}$ is a constant defined in Condition \uppercase\expandafter{\romannumeral2} for $\epsilon:=\sqrt{(-1)^{\frac{p-1}{2}}}$. 
\end{lemma}

\begin{IEEEproof}
$N_{0}(F^*)$ was determined in \cite{WangJ2}. For $N_{1}(F^*)$ and $N_{-1}(F^*)$, they can be determined in a similar way as that of Lemma \ref{lem17}. The proof is omitted here. 
\end{IEEEproof}

\subsection{The first construction of codebooks}
Let $p$ be an odd prime, $V_n^{(p)}=\gf_{p^{n_1}}\times \gf_{p^{n_2}}\times\cdots\times \gf_{p^{n_s}}$ for $n=\sum_{j=1}^{s}n_{j}$ and $F: V_{n}^{(p)}\rightarrow \gf_{p^m}$ be a vectorial dual-bent function satisfying Condition \uppercase\expandafter{\romannumeral1}.  With the same notations as in Condition \uppercase\expandafter{\romannumeral1}, we define a set 
\begin{eqnarray*}
D=\{x\in V_n^{(p)}: \eta_m(F(x))=-1\},
\end{eqnarray*}
where $K=| D|$. Define a vector of length  $K$ as
\begin{eqnarray*}
\mathbf{c}_b=\frac{1}{\sqrt{K}}(\chi_b(x))_{x\in D}.
\end{eqnarray*}
Then we obtain a $(p^n,K)$ codebook as follows:
\begin{eqnarray*}
\mathcal{C}_D=\left\{\mathbf{c}_b=\frac{1}{\sqrt{K}}(\chi_b(x))_{x\in D}: b \in V_n^{(p)}\right\},
\end{eqnarray*}
where $K=\frac{(p^n-\varepsilon p^{\frac{n}{2}})(p^m-1)}{2p^m}$ by Lemma \ref{lem17}.
\begin{theorem}\label{555}
Let $p$ be an odd prime and $F(x)$ be a vectorial dual-bent function satisfying Condition \uppercase\expandafter{\romannumeral1}.
Let the notations be the same as those in Condition \uppercase\expandafter{\romannumeral1}.
Then $\mathcal{C}_D$ is a $(p^n,K)$ asymptotically optimal codebook with the maximal cross-correlation amplitude
$$I_{\max}(\mathcal{C}_D)=\frac{p^m+1}{2p^mK} p^{\frac{n}{2}},$$
where $K=\frac{(p^n-\varepsilon p^{\frac{n}{2}})(p^m-1)}{2p^m}$ and $\varepsilon\in \{\pm 1\}$ was defined in Condition \uppercase\expandafter{\romannumeral1}.
Besides, for two distinct codewords $\mathbf{c}_{b_1}, \mathbf{c}_{b_2}\in \mathcal{C}_D$,  their cross-correlation amplitude $\mid\mathbf{c}_{b_1}\mathbf{c}_{b_2}^{H}\mid$ has the following value distribution:
\begin{eqnarray*}
\left|\mathbf{c}_{b_1}\mathbf{c}_{b_2}^{H}\right|=\left\{
\begin{array}{ll}
\frac{p^m-1}{2p^mK} p^{\frac{n}{2}},  &   \mbox{ $\frac{p^{2n}(p^m+1)}{2p^m}+\frac{\varepsilon p^{\frac{3n}{2}}(p^m-1)}{2p^m}-p^n$ times},\\
\frac{p^m+1}{2p^mK} p^{\frac{n}{2}}, &   \mbox{ $\frac{(p^{2n}-\varepsilon p^{\frac{3n}{2}})(p^m-1)}{2p^m}$ times},
\end{array} \right.
\end{eqnarray*}
where $b_1,b_2\in V_n^{(p)}$ and $b_1\neq b_2$.
\end{theorem}

\begin{IEEEproof}
We define 
\begin{eqnarray*}
\delta(x)=\left\{
\begin{array}{ll}
\frac{1-\eta_{m}(F(x))}{2},  &   \mbox{if $F(x)\neq0$},\\
0,   &   \mbox{if $F(x)=0$}.\\
\end{array} \right.
\end{eqnarray*}
For two distinct codewords $\mathbf{c}_{b_1}, \mathbf{c}_{b_2}\in \mathcal{C}_D$,  their cross-correlation is given by
\begin{eqnarray*}
\mathbf{c}_{b_1}\mathbf{c}_{b_2}^{H}&=&\frac{1}{K}\sum_{x \in D} \chi_{b_1}(x)\overline{\chi_{b_2}}(x)\\
&=&\frac{1}{K}\sum_{x \in D} \chi_{b_1-b_2}(x)\\
&=&\frac{1}{K}\sum_{x \in D} \chi_{b}(x)\\
&=&\frac{1}{K}\sum_{x \in V_{n}^{(p)}} \chi_{b}(x)\delta(x)\\
&=&\frac{1}{K}\sum_{\substack{x \in V_n^{(p)}\\F(x)\neq0}}\chi_b(x)\frac{1-\eta_m(F(x))}{2}\\
&=&\frac{1}{K}\sum_{x \in V_n^{(p)}}\chi_b(x)\frac{1-\eta_m(F(x))}{2}-\frac{1}{K}\sum_{\substack{x \in V_n^{(p)}\\F(x)=0}}\chi_b(x)\frac{1-\eta_m(F(x))}{2}\\
&=&-\frac{1}{2K}\sum_{x \in V_{n}^{(p)}}\chi_b(x)\eta_m(F(x))-\frac{1}{2K}\sum_{\substack{x \in V_n^{(p)}\\F(x)=0}}\chi_b(x),
\end{eqnarray*}
where $b:=b_1-b_2\neq0$.
According to Theorem \ref{111} and Lemma \ref{lem14}, we drive
\begin{eqnarray*}
\mathbf{c}_{b_1}\mathbf{c}_{b_2}^{H}=
\left\{
\begin{array}{ll}
-\frac{p^m-1}{2p^mK}\varepsilon p^{\frac{n}{2}}  &   \mbox{if $F^*(b)=0$ or $\eta_{m}(F^*(b))=1$},\\
\frac{p^m+1}{2p^mK}\varepsilon p^{\frac{n}{2}}   &   \mbox{if $\eta_{m}(F^*(b))=-1$}.\\
\end{array} \right.
\end{eqnarray*}
For any $b \in V_n^{(p)}\setminus \{\textbf{0}\}$, there exist $p^n$ pairs of $(b_1,b_2)$ such that $b_1-b_2=b$.
By Lemma \ref{lem19}, we then have 
\begin{eqnarray*}
\mathbf{c}_{b_1}\mathbf{c}_{b_2}^{H}=\left\{
\begin{array}{ll}
-\frac{p^m-1}{2p^mK}\varepsilon p^{\frac{n}{2}}  &   \mbox{ $\frac{p^{2n}(p^m+1)}{2p^m}+\frac{\varepsilon p^{\frac{3n}{2}}(p^m-1)}{2p^m}-p^n$ times},\\
 \frac{p^m+1}{2p^mK}\varepsilon p^{\frac{n}{2}}  &   \mbox{ $\frac{(p^{2n}-\varepsilon p^{\frac{3n}{2}})(p^m-1)}{2p^m}$ times}.\\
\end{array} \right.
\end{eqnarray*}
It is easy to deduce that
\begin{eqnarray*}
I_{\max}(\mathcal{C}_D)=\max\left\{\frac{1}{K}\left|\frac{p^m-1}{2p^m}\varepsilon p^{\frac{n}{2}}\right|,\frac{1}{K}\left|\frac{p^m+1}{2p^m}\varepsilon p^{\frac{n}{2}}\right|\right\}=\frac{p^m+1}{2p^mK} p^{\frac{n}{2}}.
\end{eqnarray*}
By Lemma \ref{welch}, it is easy to verify
\begin{eqnarray*}
\frac{I_{\max}(\mathcal{C}_D)}{I_{W}}&=&\frac{p^m+1}{2p^mK} p^{\frac{n}{2}}\sqrt{\frac{K(N-1)}{N-K}}\\
&=&\sqrt{\frac{p^{2n+2m}+2p^{2n+m}+p^{2n}-p^{n+2m}-2p^{n+m}-p^m}{p^{2n+2m}-(p^n+\varepsilon p^{\frac{n}{2}+m}-\varepsilon p^{\frac{n}{2}})^2}}\\
&\leq&\sqrt{\frac{1+\frac{1}{p^m}+\frac{1}{p^{2m}}}{1-(\frac{1}{p^m}+\frac{1}{\varepsilon p^{\frac{n}{2}}}-\frac{1}{\varepsilon p^{\frac{n}{2}+m}} )^2}}.\\
&<&\sqrt{\frac{1+\frac{2}{p^m}}{1-(\frac{2}{p^m})^2}}=\sqrt{\frac{1}{1-\frac{2}{p^m}}},
\end{eqnarray*}
where $n\geq \frac{m}{2}$.
Thus $\frac{I_{\max}(\mathcal{C}_D)}{I_{W}}\rightarrow 1$ if $p\rightarrow+\infty$ or $m\rightarrow +\infty$.
Then the desired conclusions follow. 
\end{IEEEproof}

\begin{table}
\caption{Asymptotically optimal codebooks in Theorem \ref{555}}\label{tab1}
\centering
\begin{tabular}{cccccc}
   \toprule
    $p$&$N$ & $K$ & $I_{\max}$ & $I_{W}$ & $I_{W}/I_{\max}$\\
   \midrule
   3&9   & 4  & 0.5000 & 0.3954 & 0.7908  \\ 
   3&6561 & 2880 & 0.015625 & 0.0140 & 0.8960  \\
   3&531441&255528& 0.00147928&  0.00142541  &    0.9635\\  
   3& 43046721  & 21254400  & 0.00015625 & 0.00015433& 0.9877  \\ 
   3&3486784401   & 1736188344  & 0.000017075 & 0.000017005 & 0.9959  \\                  
   \bottomrule
\end{tabular}
\end{table}

\begin{remark}
By Theorem \ref{555},  $\frac{I_{\max}(\mathcal{C}_D)}{I_{W}}\rightarrow 1$ if $p\rightarrow+\infty$ or $m\rightarrow +\infty$.
If we fix a very small $p$ and let $m\rightarrow +\infty$, then the parameters of the codebook $\mathcal{C}_D$ in Theorem \ref{555} 
also asymptotically achieve the Welch bound. Hence the codebook  $\mathcal{C}_D$ could have both very small alphabet size and small maximal cross-correlation amplitude. 
For instance, in Table \ref{tab1}, if we fix $p=3$ and choose some small $m$, then we derive the parameters of the corresponding codebooks by Theorem \ref{555}. 
The numerical data show that the codebook $\mathcal{C}_D$ in Theorem \ref{555} is indeed asymptotically optimal with respect to the Welch bound.

When $\varepsilon=-1$, the parameters of $\mathcal{C}_D$ in Theorem $\ref{555}$ are the same as those in \cite{WuX}.
However, if $\varepsilon=1$, then $\mathcal{C}_D$ in Theorem $\ref{555}$ has different parameters. 
Note that our construction of codebooks works for all vectorial dual-bent function 
satisfying Condition \uppercase\expandafter{\romannumeral1} and contains the construction in \cite{WuX} as a spacial case. 
\end{remark}

\begin{example}
Let $p=3, m=t=2,n=8$, $\omega$ be a primitive element of $\gf_{3^8}$ and $F(x)=\tr_{3^8/3^2}(\omega x^2)$. By Equation (\ref{Tr(x^2)}), $F(x)$ is a vectorial dual-bent function satisfying Condition \uppercase\expandafter{\romannumeral1} with $l=d=2$ and $\varepsilon=1$. The $\mathcal{C}_D$ is a $(3^8, 2880)$ codebook with $I_{\max}(\mathcal{C}_D)=\frac{1}{64}$.
\end{example}

\subsection{A pair of sequences and the second construction of codebooks}
In this subsection, we study the cross-correlation of a pair of sequences and give the second construction of complex codebooks by applying the hybrid character sum $S_1$.

\subsubsection{The cross-correlation of a pair of sequences}
To begin with, we recall the following quadratic residue mapping as follows.
 \begin{Definition}\cite{HongS}
The quadratic residue mapping $q_{m}:\gf_{p^m}\rightarrow \gf_2$ is defined as
\begin{eqnarray*}
q_m(x)=\left\{
\begin{array}{ll}
1  &   \mbox{if $\eta_m(x)=-1$},\\
0    &   \mbox{if $\eta_m(x)=1$ or $x=0$}.
\end{array} \right.
\end{eqnarray*}
 \end{Definition}

In the following, we study the cross-correlation of a pair of sequences by using the hybrid character sum $S_1$.
Let $p$ be an odd prime, $V_n^{(p)}=\gf_{p^n}$ and $F(x)$ be a vectorial dual-bent function  satisfying Condition \uppercase\expandafter{\romannumeral1}. 
This pair of sequences consists of the $p$-ary $m$-sequence 
$$\textbf{s}_1:=(\tr_{p^n/p}(\alpha^t))_{t=0}^{p^n-2}$$ and the binary sequence
$$\textbf{s}_2:=\left(q_m(F(\alpha^t)) \right)_{t=0}^{p^n-2},$$
where $\gf_{p^n}^*=\langle\alpha\rangle$.
Their cross-correlation is defined by
\begin{eqnarray*}
C(\tau)
&=&\sum_{t=0}^{p^n-2}\zeta_{p}^{\tr_{p^n/p}(\alpha^{t+\tau})}\cdot(-1)^{q_m(F(\alpha^t))}\\
&=&\sum_{x \in \gf_{p^n}^{*}}\chi_{1}(ax)\cdot(-1)^{q_{m}(F(x))},
\end{eqnarray*}
where $\tau=0,1,\cdots, p^n-2$ and $a:=\alpha^{\tau} \in \gf_{p^n}^*$.

\begin{theorem}\label{fenb}
Let $p$ be an odd prime, $V_n^{(p)}=\gf_{p^n}$ and $F(x)$ be a vectorial dual-bent function  satisfying Condition \uppercase\expandafter{\romannumeral1}. 
If $\tau$ runs over the set $\{0,1,\cdots, p^n-2\}$, then $C(\tau)$ has the following distribution:
\begin{eqnarray*}
C(\tau)=\left\{
\begin{array}{ll}
\frac{p^m-1}{p^m}\varepsilon p^{\frac{n}{2}}-1  &   \mbox{ $-1+\frac{(p^m+1)p^n+\varepsilon p^{\frac{n}{2}}(p^m-1)}{2p^m}$ times},\\
-\frac{p^m+1}{p^m}\varepsilon p^{\frac{n}{2}}-1   &   \mbox{ $\frac{(p^n-\varepsilon p^{\frac{n}{2}})(p^m-1)}{2p^m}$ times},
\end{array} \right.
\end{eqnarray*}
where  $\varepsilon\in \{\pm 1\}$ was defined in Condition \uppercase\expandafter{\romannumeral1}.
\end{theorem}

\begin{IEEEproof}
Note that
\begin{eqnarray*}
C(\tau)
&=&\sum_{x \in \gf_{p^n}^{*}}\chi_{1}(ax)\cdot(-1)^{q_{m}(F(x))}\\
&=&\sum_{x \in \gf_{p^n}^{*}}\chi_{1}(ax)\eta_{m}(F(x))+\sum_{\substack{x \in \gf_{p^n}^*\\F(x)=0}}\chi_a(x)\\
&=&S_1+\sum_{\substack{x \in \gf_{p^n}^*\\F(x)=0}}\chi_a(x).
\end{eqnarray*}
By Theorem \ref{111} and Lemma \ref{lem14}, we have 
\begin{eqnarray*}
C(\tau)=\left\{
\begin{array}{ll}
\frac{p^m-1}{p^m}\varepsilon p^{\frac{n}{2}}-1  &   \mbox{if $F^*(a)=0$ or $\eta_m(F^*(a))=1$},\\
-\frac{p^m+1}{p^m}\varepsilon p^{\frac{n}{2}}-1   &   \mbox{if $\eta_m(F^*(a))=-1$}.
\end{array} \right.
\end{eqnarray*}
By Lemma \ref{lem17}, it is easy to obtain the frequency of each value of $C(\tau)$.
\end{IEEEproof}

\begin{remark}
Let $|C(\tau)|_{\max}$ denote the maximum cross-correlation magnitude of $\textbf{s}_1$ and $\textbf{s}_2$. 
By Theorem \ref{fenb}, if $m\rightarrow +\infty$, then $|C(\tau)|_{\max}\approx p^{\frac{n}{2}}+1$, which indicates that the pair of sequences $\textbf{s}_1$ and $\textbf{s}_2$
has low maximum cross-correlation magnitude.
\end{remark}

\subsubsection{The second construction of codebooks}
Let $p$ be an odd prime, $V_n^{(p)}=\gf_{p^n}$ and $F(x)$ be a vectorial dual-bent function  satisfying Condition \uppercase\expandafter{\romannumeral1}. In the following, we present a construction of partial Hadamard codebooks.To this end, we define a row selection sequence as follows. 

\begin{Definition}\label{def3}
Define the binary row selection sequence
\begin{eqnarray*}
\mathbf{r}_m=\{r_0,r_1,\cdots,r_{p^n-1}\}
\end{eqnarray*}
by
\begin{eqnarray*}
r_k=\left\{
\begin{array}{ll}
q_m(F(\alpha^{k-1}))  &   \mbox{for $1\leq k \leq p^n-1$},\\
1    &   \mbox{for $k=0$}.
\end{array} \right.
\end{eqnarray*}
\end{Definition}

 Let $\mathbf{H}=[h_{i,j}]_{0\leq i,j\leq p^n-1}$ be the $p$-ary Hadamard matrix in Definition \ref{def1}. Let $N=p^n$ and $K=\frac{(p^n-\varepsilon p^{\frac{n}{2}})(p^m-1)}{2p^m}+1$. Let $D=\{d_0,d_1,\cdots,d_{K-1}\}$ be the support of the binary row selection sequence $\mathbf{r}_m$ in Definition \ref{def3}, where $| D|=\frac{(p^n-\varepsilon p^{\frac{n}{2}})(p^m-1)}{2p^m}+1=K$. Let $\mathcal{C}_{H}(\mathbf{r}_m)=\{\mathbf{c}_0,\mathbf{c}_1,\ldots,\mathbf{c}_{N-1}\}$ be the $(N,K)$ partial Hadamard codebook whose code vectors are the columns of the submatrix obtianed by selecting $K$ rows corresponding to $D$ from the $p^n\times p^n$ Hadamard matrix $\mathbf{H}=[h_{i,j}]_{0\leq i,j\leq p^n-1}$.

\begin{theorem}\label{777}
Let $p$ be an odd prime, $V_n^{(p)}=\gf_{p^n}$ and $F(x)$ be a vectorial dual-bent function  satisfying Condition \uppercase\expandafter{\romannumeral1}. Then $\mathcal{C}_{H}(\textbf{r}_m)$ is an $(N,K)$ asymptotically optimal partial Hadamard codebook with \begin{eqnarray*}
I_{\max}(\mathcal{C}_{H}(\mathbf{r}_m))=\max \left\{\frac{1}{2K}\left| -\frac{p^m+1}{p^m}\varepsilon p^{\frac{n}{2}}-2 \right|, \frac{1}{2K}\left| \frac{p^m-1}{p^m}\varepsilon p^{\frac{n}{2}}-2 \right| \right\},
\end{eqnarray*} where $N=p^n$ and $K=\frac{(p^n-\varepsilon p^{\frac{n}{2}})(p^m-1)}{2p^m}+1$. Let $I_{i,j}(\mathcal{C}_{H}(\textbf{r}_m))$ be the magnitude of the inner product between the code vectors $\mathbf{c}_i$ and $\mathbf{c}_j$, where $i, j$ are different integers and $0\leq i, j\leq p^n-1$. Then $I_{i,j}(\mathcal{C}_{H}(\mathbf{r}_m))$ has the following distribution:
\begin{eqnarray*}
I_{i,j}(\mathcal{C}_{H}(r_m))=\left\{
\begin{array}{ll}
\frac{1}{2K}\left|\frac{p^m-1}{p^m}\varepsilon p^{\frac{n}{2}}-2\right|  &   \mbox{ $-p^n+\frac{(p^m+1)p^{2n}+\varepsilon p^{\frac{3n}{2}}(p^m-1)}{2p^m}$ times},\\
\frac{1}{2K}\left|-\frac{p^m+1}{p^m}\varepsilon p^{\frac{n}{2}}-2\right|   &   \mbox{ $\frac{p^n(p^n-\varepsilon p^{\frac{n}{2}})(p^m-1)}{2p^m}$ times},
\end{array} \right.
\end{eqnarray*}
where  $\varepsilon\in \{\pm 1\}$ was defined in Condition \uppercase\expandafter{\romannumeral1}.
\end{theorem}
\begin{IEEEproof}
 Let $\mathbf{r}_m=\{r_0,r_1,\cdots,r_{p^n-1}\}$ be the binary row selection sequence in Definition \ref{def3}. By Lemma \ref{bina}, there exist $p^n$ distinct pairs of $(i,j)$ satisfying 
 \begin{eqnarray*}
 I_{i,j}(\mathcal{C}_{H}(\mathbf{r}_m))
 &=&\frac{1}{2K}|\widehat{r}_l|\\
  &=&\frac{1}{2K}\left|\sum_{k=0}^{p^n-1}(-1)^{r_k}h_{k,l}\right|\\
  &=&\frac{1}{2K}\left|-1+\sum_{k=1}^{p^n-1}(-1)^{q_m(F(\alpha^{k-1}))}\zeta_{p}^{\tr_{p^n/p}(\alpha^{k+l-2})}\right|\\
 &=&\frac{1}{2K}\left|-1+\sum_{k=0}^{p^n-2}(-1)^{q_m(F(\alpha^{k}))}\zeta_{p}^{\tr_{p^n/p}(\alpha^{k+l-1})}\right|\\ 
 &=&\frac{1}{2K}\left| -1+ C(l-1)\right|.
 \end{eqnarray*}
By Theorem \ref{fenb}, it is easy to obtain the value distribution of $I_{i,j}(\mathcal{C}_{H}(\mathbf{r}_m))$, where  $i, j$ are different integers and $0\leq i, j\leq p^n-1$.
It is obvious that
  \begin{eqnarray*}
I_{\max}(\mathcal{C}_{H}(\mathbf{r}_m))
&=&\max \left\{\frac{1}{2K}\left| -\frac{p^m+1}{p^m}\varepsilon p^{\frac{n}{2}}-2 \right|, \frac{1}{2K}\left| \frac{p^m-1}{p^m}\varepsilon p^{\frac{n}{2}}-2 \right| \right\}\\
 &\leq& \frac{1}{2K}\left(\frac{p^m+1}{p^m} p^{\frac{n}{2}}+2\right).
  \end{eqnarray*}
  By the Welch bound, we have
  \begin{eqnarray*}
\frac{I_{\max}(\mathcal{C}_{H}(\mathbf{r}_m))}{I_{W}}
&\leq&\frac{1}{K}\left(\frac{p^m+1}{2p^m} p^{\frac{n}{2}}+1\right)\sqrt{\frac{K(N-1)}{N-K}}\\
&=&\left(\frac{p^m+1}{2p^m} p^{\frac{n}{2}}+1\right)\sqrt{\frac{4p^{n+2m}-4p^{2m}}{p^{2n+2m}-(p^n+\varepsilon p^{\frac{n}{2}+m}-\varepsilon p^{\frac{n}{2}}-p^m)^2}}\\
&\leq&\sqrt{\frac{1+\frac{2}{p^m}+\frac{4}{p^{2m}}}{1-\frac{(p^n+\varepsilon p^{\frac{n}{2}+m}-\varepsilon p^{\frac{n}{2}}-p^m)^2}{p^{2n+2m}}}}\\
&<&\sqrt{\frac{1+\frac{4}{p^m}}{1-\frac{16}{p^{2m}}}}=\sqrt{\frac{1}{1-\frac{4}{p^m}}}.
\end{eqnarray*}
Thus $\frac{I_{\max}(\mathcal{C}_{H}(\mathbf{r}_m))}{I_{W}}\rightarrow 1$ if $p\rightarrow+\infty$ or $m\rightarrow +\infty$.
\end{IEEEproof}

 \begin{remark}
 If we fix a very small $p$ and let $m\rightarrow +\infty$, then the parameters of the codebook $\mathcal{C}_{H}(\textbf{r}_m)$  in Theorem \ref{777} 
 asymptotically achieve the Welch bound. Hence the codebook  $\mathcal{C}_{H}(\textbf{r}_m)$ could have both very small alphabet size and small maximal cross-correlation amplitude. 
 Compared to the codebook in \cite[Theorem 34]{Heng}, our codebook in Theorem \ref{777} has more abundant and flexible parameters. 
 In fact, we can obtain the codebook in \cite[Theorem 34]{Heng} from Theorem \ref{777} if $m=1$.
 \end{remark}

\subsection{The third construction of codebooks}
Let $p$ be an odd prime, $V_n^{(p)}=\gf_{p^{n_1}}\times \gf_{p^{n_2}}\times\cdots\times \gf_{p^{n_s}}$ and $F: V_{n}^{(p)}\rightarrow \gf_{p^m}$ be a vectorial dual-bent function satisfying Condition \uppercase\expandafter{\romannumeral2}. Define a set 
\begin{eqnarray*}
D_1=\left\{x\in V_n^{(p)} \setminus \{0\}: F(x)=0\right\},
\end{eqnarray*}
where $K=| D_1|$.
We define a codebook of length  $K$ as
\begin{eqnarray*}
\mathbf{c}_b=\frac{1}{\sqrt{K}}(\chi_b(x))_{x\in D_1}.
\end{eqnarray*}
Then an $(N,K)$ codebook is given as
\begin{eqnarray*}
\mathcal{C}_{D_1}=\left\{\mathbf{c}_b=\frac{1}{\sqrt{K}}(\chi_b(x))_{x\in D_1}: b \in V_n^{(p)}\right\}.
\end{eqnarray*}
It is easy to deduce that $N=p^{n}$ and $K=p^{n-m}-1$ by Lemma \ref{lem16}.

\begin{theorem}\label{666}
Let $p$ be an odd prime and $F(x)$ be a vectorial dual-bent function satisfying Condition \uppercase\expandafter{\romannumeral2}.
Let the notations be the same as those in Condition \uppercase\expandafter{\romannumeral2}.
Then $\mathcal{C}_{D_1}$ is an $(p^{n},p^{n-m}-1)$ asymptotically optimal codebook with
\begin{eqnarray*}
I_{\max}(\mathcal{C}_{D_1})=\max \left\{\frac{1}{K}\left|\frac{1}{p^m}\upsilon p^{\frac{n}{2}}(-1)^{m-1}\epsilon^m\sqrt{p^m}-1 \right|, \frac{1}{K}\left|\frac{1}{p^m}\upsilon p^{\frac{n}{2}}(-1)^{m-1}\epsilon^m\sqrt{p^m}+1\right| \right\},
\end{eqnarray*}
where $K=p^{n-m}-1$ and $\upsilon \in \{\pm\epsilon^{m}\}$ is a constant defined in Condition \uppercase\expandafter{\romannumeral2} for $\epsilon:=\sqrt{(-1)^{\frac{p-1}{2}}}$. 
Let $\mathbf{c}_{d_1}, \mathbf{c}_{d_2}$ be two codewords in $\mathcal{C}_{D_1}$ for $d_1\neq d_2$ and $d_1,d_2\in V_n^{(p)}$. Then $|\mathbf{c}_{d_1}\mathbf{c}_{d_2}^{H}|$ has the following value distribution:
\begin{eqnarray*}
\left|\mathbf{c}_{d_1}\mathbf{c}_{d_2}^{H}\right|=\left\{
\begin{array}{lll}
\frac{1}{K} &   \mbox{ $p^{2n-m}-p^{n}$ times},\\
\frac{1}{K}\left|\frac{1}{p^m}\upsilon p^{\frac{n}{2}}(-1)^{m-1}\epsilon^m\sqrt{p^m}-1\right| &   \mbox{ $p^n\frac{(p^{n-m}+(-1)^{m-1}\epsilon^m\upsilon^{-1}\eta_m(-1) p^{\frac{n-m}{2}})(p^m-1)}{2}$ times},\\
\frac{1}{K}\left|\frac{1}{p^m}\upsilon p^{\frac{n}{2}}(-1)^{m-1}\epsilon^m\sqrt{p^m}+1\right|   &   \mbox{ $p^n\frac{(p^{n-m}-(-1)^{m-1}\epsilon^m\upsilon^{-1}\eta_m(-1) p^{\frac{n-m}{2}})(p^m-1)}{2}$ times}.\\
\end{array} \right.
\end{eqnarray*}
\end{theorem}

\begin{IEEEproof}
Let $\mathbf{c}_{d_1}, \mathbf{c}_{d_2}$ be two codewords in $\mathcal{C}_{D_1}$ for $d_1\neq d_2$ and $d_1,d_2\in V_n^{(p)}$. Then
\begin{eqnarray*}
\mathbf{c}_{d_1}\mathbf{c}_{d_2}^{H}&=&\frac{1}{K}\sum_{x \in D_1} \chi_{d_1}(x)\overline{\chi_{d_2}}(x)\\
&=&\frac{1}{K}\sum_{x \in D_1} \chi_{d_1-d_2}(x)\\
&=&\frac{1}{K}\sum_{\substack{x \in V_n^{(p)}\setminus \{0\} \\F(x)=0}}\chi_{d}(x),
\end{eqnarray*}
where $d:=d_1-d_2\neq0$.
By Lemma\ref{lem15}, we have
\begin{eqnarray*}
K(\mathbf{c}_{d_1}\mathbf{c}_{d_2}^{H})&=&\left\{
\begin{array}{lll}
-1  &   \mbox{if $F^*(d)=0$},\\
\frac{1}{p^m}\upsilon p^{\frac{n}{2}}(-1)^{m-1}\epsilon^m\sqrt{p^m}-1  &   \mbox{if $\eta_m(F^*(d))=1$},\\
-\frac{1}{p^m}\upsilon p^{\frac{n}{2}}(-1)^{m-1}\epsilon^m\sqrt{p^m}-1   &   \mbox{if $\eta_m(F^*(d))=-1$}.\\
\end{array} \right.
\end{eqnarray*}
Note that for any $d \in V_n^{(p)}\backslash \{0\}$, there exist $p^n$ pairs of $(d_1,d_2)$ such that $d_1-d_2=d$.
According to Lemma \ref{lem20}, we then have 
\begin{eqnarray*}
\mathbf{c}_{d_i}\mathbf{c}_{d_j}^{H}=\left\{
\begin{array}{lll}
-\frac{1}{K} &   \mbox{ $p^{2n-m}-p^{n}$ times},\\
\frac{1}{K}\left(\frac{1}{p^m}\upsilon p^{\frac{n}{2}}(-1)^{m-1}\epsilon^m\sqrt{p^m}-1\right) &   \mbox{ $p^n\frac{(p^{n-m}+(-1)^{m-1}\epsilon^m\upsilon^{-1}\eta_m(-1) p^{\frac{n-m}{2}})(p^m-1)}{2}$ times},\\
-\frac{1}{K}\left(\frac{1}{p^m}\upsilon p^{\frac{n}{2}}(-1)^{m-1}\epsilon^m\sqrt{p^m}+1\right)   &   \mbox{ $p^n\frac{(p^{n-m}-(-1)^{m-1}\epsilon^m\upsilon^{-1}\eta_m(-1) p^{\frac{n-m}{2}})(p^m-1)}{2}$ times}.\\
\end{array} \right.
\end{eqnarray*}
Then the maximum cross-correlation magnitude follows. It is obvious that
\begin{eqnarray*}
I_{\max}(\mathcal{C}_{D_1})\leq\frac{1}{K}\left(p^{\frac{n-m}{2}}+1\right).
\end{eqnarray*}
By Lemma \ref{welch}, it is easy to verify
\begin{eqnarray*}
\frac{I_{\max}(\mathcal{C}_{D_1})}{I_{W}}\leq\frac{1}{K}(p^{\frac{n-m}{2}}+1)\sqrt{\frac{K(N-1)}{N-K}}<\sqrt{\frac{1}{1-\frac{1}{p^m}}}.
\end{eqnarray*}
Thus $\frac{I_{\max}(\mathcal{C}_{D_1})}{I_{W}}\rightarrow 1$ if $p\rightarrow+\infty$ or $m\rightarrow +\infty$.
\end{IEEEproof}

\begin{table}
\caption{Asymptotically optimal codebooks in Theorem \ref{666}}\label{tab-2}
\centering
\begin{tabular}{cccccc}
   \toprule
    $p$&$N$ & $K$ & $I_{\max}$ & $I_{W}$ & $I_{W}/I_{\max}$\\
   \midrule
  3& 243  & 80 & 0.125 & 0.09176 & 0.73406 \\ 
   3& 729  & 80  & 0.125 & 0.10556 & 0.8445 \\
   3&19683&728&0.03846&0.036337&0.94476\\
   3&531441 & 6560 & 0.01250 & 0.012269 & 0.981615 \\
  3& 14348907  & 59048 & 0.00413223 & 0.0041068 & 0.99384234 \\                            
   \bottomrule
\end{tabular}
\end{table}

\begin{remark}
By Theorem \ref{666}, $\frac{I_{\max}(\mathcal{C}_{D_1})}{I_{W}}\rightarrow 1$ if $p\rightarrow+\infty$ or $m\rightarrow +\infty$.
If we fix a very small $p$ and let $m\rightarrow +\infty$, then the parameters of the codebook $\mathcal{C}_{D_1}$ in Theorem \ref{666} 
asymptotically achieve the Welch bound. Hence the codebook  $\mathcal{C}_{D_1}$ could have both very small alphabet size and small maximal cross-correlation amplitude. 
For instance, in Table \ref{tab-2}, if we fix $p=3$ and choose some small $m$, then we derive the parameters of the corresponding codebooks by Theorem \ref{666}. 
The numerical data show that the codebook $\mathcal{C}_{D_1}$ in Theorem \ref{666} is indeed asymptotically optimal with respect to the Welch bound.
\end{remark}

\begin{example}
Let $p=3, m=t=2,n=6$, $\omega_1$ be a primitive element of $\gf_{3^6}$ and $F(x)=\tr_{3^6/3^2}(\omega_1 x^2)$. Then by Equation (\ref{Tr(x^2)1}), $F(x)$ is a vectorial dual-bent function satisfying Condition \uppercase\expandafter{\romannumeral2} with $l=d=2$ and $\upsilon=-1$. Then $\mathcal{C}_{D_1}$ is a $(3^6, 80)$ codebook with $I_{\max}(\mathcal{C}_{D_1})=\frac{1}{8}$.
\end{example}

\section{Concluding reamrks}
In this paper, we generalized the hybrid character sums from the form (\ref{eqn-1}) to the form (\ref{form2}).
By Gaussian sums, we studied the absolute values or explicit values of the hybrid character sums in form (\ref{form2}) if $F(x)$ is a vectorial dual-bent function satisfying Condition \uppercase\expandafter{\romannumeral1} or \uppercase\expandafter{\romannumeral2} and $G(x)=ax$.
It is remarkable that the hybrid character sums studied by us have small absolute values, which indicates that they have potential applications in coding theory and sequence designs. 
As applications, we presented three constructions of asymptotically optimal codebooks whose maximal cross-correlation amplitudes were determined by the hybrid character sums.
Each family of these codebooks has small alphabet size $p$, which enhances their appeal for implementation. 
 Compared with the codebooks in \cite{Heng} \cite{WuX}, the codebooks constructed in this paper have more abundant and flexible parameters.

\end{document}